\documentclass[11pt, letter,fleqn]{article}
\pdfoutput=1

\usepackage{hyperref}
\usepackage{amsmath, amssymb,amsthm,cite}
\usepackage{graphics,color}
\usepackage{subcaption}
\usepackage{cprotect}
\usepackage{epstopdf}
\usepackage {graphicx}
\usepackage{verbatim}
\usepackage{listings}
\usepackage{tikz}
\usepackage[section]{placeins}
\usepackage{multirow}
\usepackage{longtable}	%pagebreaking for tables
\usepackage{pdflscape}	
\usepackage{enumitem}  %for nested itemize
\usepackage{placeins} %use for float barrier

\usepackage{multicol}
\usepackage{matlab-prettifier}
\usepackage{tabularx}

\title{A Method for Random Packing of Spheres with Application to Bonding Modeling in Powder Bed 3D Printing Process }

\author{  Travis J. Black\footnotemark[1], ~~Alexei
  F. Cheviakov\footnotemark[2] \\ {\small
    \emph{Department of Mathematics and Statistics, University of
      Saskatchewan, Saskatoon, S7N 5E6 Canada}}
}

 {\theoremstyle{definition}

\setlength{\parskip}{.5ex}

{\theoremstyle{definition} }
{\theoremstyle{definition} }

\newcommand{\vspacebefore}{\raisebox{0ex}[2.5ex][0ex]{\null}}

\def\beq{\begin{equation}}
\def\eeq{\end{equation}}
\def\barr{\begin{array}{ll}}
\def\earr{\end{array}}

\begin{document}

\footnotetext[1]{Corresponding author. Electronic mail: tjb235@usask.ca
  }
%\footnotetext[2]{Work done within NSERC Undergraduate Summer Research
%Fellowship.}
\footnotetext[2]{Electronic mail: cheviakov@math.usask.ca}

\maketitle

\begin{abstract}

A \verb|Matlab|-based computational procedure is proposed to fill a given volume with spheres whose radii are randomly picked from any specified probability distribution supported by \verb|Matlab|. The general program sequence and examples of filling a unit cube, a parallelepiped, and a concave domain between two hemispherical surfaces, with spheres whose radii are drawn from the Weibull and Gamma distributions, are presented. A sample application to the numerical modeling of bond formation between particles heated by a laser beam in powder bed 3D printing process is considered.

\end{abstract}
%\begin{keyword}
%
%\end{keyword}
%\end{frontmatter}
%

{\bf PROGRAM SUMMARY}
  %Delete as appropriate.

\begin{small}
\noindent
{\em Manuscript Title:} A Method for Random Packing of Spheres with Application to Bonding Modeling in Powder Bed 3D Printing Process  \\
{\em Authors:} Travis J. Black and Alexei F. Cheviakov  \\
{\em Program Title:}  RSPP Random Sphere Packing Program\\
{\em Journal Reference:}      \\
  %Leave blank, supplied by Elsevier.
{\em Catalogue identifier:}     \\
  %Leave blank, supplied by Elsevier.
{\em Licensing provisions:} none      \\
  %enter "none" if CPC non-profit use license is sufficient.
{\em Programming language:} Matlab 2021a     \\
{\em Computer:} PC        \\
  %Computer(s) for which program has been designed.
{\em Operating system:} Windows 10 or any other OS supporting Matlab  \\
  %Operating system(s) for which program has been designed.
{\em RAM:} { 128 \, GB (workstation configuration). Also tested with 16 GB (laptop configuration). }     \\
  %RAM in bytes required to execute program with typical data.
{\em Number of processors used:} 2 Xeon processors, 32 logical processors (workstation configuration). Also tested on an Intel i7-based laptop with 1 physical, 4 logical processors.  \\
  %If more than one processor.
  % Please give some freely chosen keywords that we can use in a
  % cumulative keyword index.
{\em Classification:}   \\
  %Classify using CPC Program Library Subject Index, see (
  % http://cpc.cs.qub.ac.uk/subjectIndex/SUBJECT_index.html)
  %e.g. 4.4 Feynman diagrams, 5 Computer Algebra.
{\em Nature of problem:}  Fill a given three-dimensional domain with a prescribed random distribution of spheres.\\
  %Describe the nature of the problem here.
{\em Solution method:} A parallelepiped-shaped (unit brick) subset of the total volume is filled with the spheres using a geometry-based approach. The unit brick is used as a building block to fill the remaining space. \\
  %Describe the method solution here.
{\em Restrictions:} In the direct application of the program, the domain must be constructed from unit bricks. More complex domains may be treated in a similar manner, with an example presented.\\
  %Describe any restrictions on the complexity of the problem here.
{\em Unusual features:} Geometry-based approach. \\
  %Describe any unusual features of the program/problem here.
{\em Running time:}
  %Give an indication of the typical running time here.
  from two minutes to 190 minutes (workstation configuration). From five minutes to 12 hours (laptop configuration).
   \\
\end{small}

\section{Introduction}

The proposed \verb|Matlab|-based code fills a given three-dimensional domain with spheres of radii following any prescribed random distribution of sphere available in \verb|Matlab|. The motivation behind the creation of this code was to build a Discrete Element Method (DEM) model of the powder bed 3D printing process. The initial step in such a model was the determination of the initial location of a packing of individual particles within the powder bed domain, and which particles were in contact. This posed a problem to which no simple solution was readily available. Indeed, finding an optimal packing of unequal spheres is a challenging task \cite{WangJie1999PoUS}. Previous DEM models (e.g., Refs.~\cite{michopoulos2018multiphysics,XIN2018373}) relied on the particle dynamics approach to determine location with gravitational and inter-particle forces; the interaction and final positions of the spheres was determined using Newtonian mechanics. This required particle-particle and particle-boundary collision detection, as well as calculation of contact forces \cite{michopoulos2018multiphysics}. For a large number of particles, such calculations can be highly computationally expensive. Since our only concern was the final placement on the particles, the current code was designed to bypass the dynamics of powder settling and find a way to determine the rest location of the particles based on geometrical considerations.

Random sphere packing has broad applications, including DEM modelling, granular dynamics, radiosurgery for treating brain tumors \cite{WangJie1999PoUS}, optimal packing problems, etc. Other sphere packing methods have similar aims \cite{HIFI2019482,StoyanYu.2016PUSi}, however they did not meet our needs. Vast literature is dedicated to a related but different problem of random packing of equally-sized spheres (see, e.g., references \cite{torquato2000random, williams2003random} and references therein).

The goal for the program presented below was not to find the \emph{optimal} packing, but rather a packing of unequal spheres which closely models a realistic packing of metal powder particles in the powder bed printing process. In the simplest setting where the total domain to be filled can be represented as a union of parallelepiped-shaped bricks, initially, a brick is filled with the distribution of spheres, and then is used as a building block to achieve the desired volume. The symmetry of the brick is exploited to keep track of particle contacts.

Two related methods were developed, and are available to the user within the current \verb|Matlab|-based package. Each consists of a part responsible for filling a single brick and a part that builds the volume.

For Method 1, the approach to filling the unit brick, an arbitrary parallelepiped-shaped volume, the main sphere fitting function, a typical program sequence and three run examples are discussed in Section \ref{sec:FullMethod1}. The first method randomly fills the edges of the unit brick, then the faces, and then the volume, which results in a unit brick filled with a non-symmetric spherical distribution. The total domain may coincide with a single brick, or be made of several bricks in $x$, $y$, and $z$ directions; in the latter case, the total domain is constructed by reflections of the unit brick about its faces, providing boundary sphere contacts.

For the second method, details and examples are provided in Section \ref{sec:FullMethod2}. In Method 2, unlike Method 1, for the unit brick, only one edge is filled in each direction, and four parallel copies of each are made. Similarly, only three faces in each plane are filled with random spheres, and are copied onto the opposite ones. This yields a unit brick that has identical opposite faces (but a non-symmetric volume filling), and hence a direct copy-paste of unit bricks can be used to fill a larger total domain.
Another difference of Method 2 from Method 1 is that instead of being fully inside of the unit brick and touching the brick faces, in Method 2, centers of the spheres on each face are located on the brick faces themselves.

Example 3 (Section \ref{sec:eg3:hemisph}) illustrates an application of the geometric approach to create a spherical filling of a more complex-shaped domain: a parallelepiped with the subtraction of two hemispheres centered in the middles of two opposite faces in the $x$-direction.

A physical example containing a simple model considering discrete laser-induced heat-based bonding in powder bed 3D printing process is considered in Section \ref{egPhys}, following Ref.~\cite{spierings2009comparison}. The physical principles and constants are described in Section \ref{egPhys:physics}, and a result of a simulated print of a small square are presented in Section \ref{egPhys:sim}.

In examples used in the current work, Weibull and Gamma probability distribution of spherical radii \cite{SteubenJohnC2016Demo,spierings2011influence} were employed. The Weibull and Gamma distribution parameters were chosen to correspond to powder bed additive manufacturing involving steel spheres. The presented software supports all probability distributions provided in \verb|Matlab|.

The paper is concluded with a summary discussion in Section \ref{sec:Discuss}.

%==================
\section{Filling a domain with spheres: Method 1} \label{sec:FullMethod1}

The first method, as well as the second method method described later, can be used to fill any parallelepiped-shaped domain $\mathcal{V}$ with a given random distribution of spheres.

The domain $\mathcal{V}$ can be filled either in a completely random manner, or for a quicker computation, it can be subdivided into smaller standard parallelepipeds (``unit bricks"). In the latter case, a single brick would be filled with spheres randomly, and bricks can be copied and joined, as explained below, any prescribed number of times in $x$, $y$, and $z$ directions, to create a filling of $\mathcal{V}$. In the former case, the full domain $\mathcal{V}$ is treated as a single unit brick.

During the spherical filling, positions and radii of random spheres are recorded, as well as pairwise connections between touching spheres, and sets of sphere indices corresponding to spheres lying on each face of $\mathcal{V}$.

\subsection{Method 1: filling the unit brick} \label{sec:FullMethod1:unitcube}

\begin{figure}
  \centering
  \includegraphics[width=0.7\textwidth]{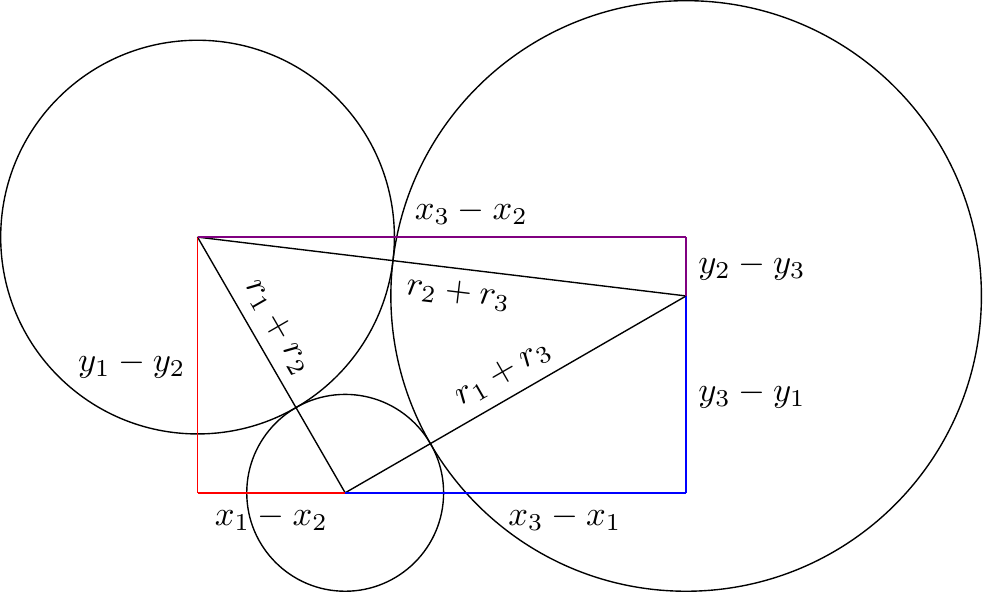}
  \caption{New sphere in contact with parent spheres: a 2D cartoon}\label{System of Equations Pic}
\end{figure}

The initial step in the first method of filling up a unit brick with a given distribution of spheres consists in placing a sphere of average radius in each corner of the brick, then filling the edges between each adjacent corners with contacting spheres with sizes drawn randomly from the same distribution, and then filling faces the same way. After all faces are finished, the remaining volume is filled. When a new sphere is placed, neighbouring spheres in contact with the new one are recorded. This way, when the volume is filled, there is a list containing all pairs of spheres that are in contact. As new spheres are placed, a list of other spheres that are in contact with the given one is kept, based on a constant dimensionless parameter $\varepsilon$ that specifies acceptable separation/overlap of two particles to be considered in contact. Particles that are considered ``close" (controlled by another constant dimensionless parameter $\delta$) are stored as possible ``parents." When a new sphere needs to be placed, to determine its location, the program runs through the list of possible ``parents," as explained below, and the new sphere is placed to be in contact with possible parent particles (while we work in 3D, the idea is shown in 2D in Figure \ref{System of Equations Pic}). This is achieved by solving a system of equations
\[
||\mathbf{x}_n - \mathbf{x}_1 || = R_n + R_1, \qquad
|| \mathbf{x}_n - \mathbf{x}_2 || = R_n + R_2, \qquad
|| \mathbf{x}_n - \mathbf{x}_3 || = R_n + R_3,
\]
where $\mathbf{x}_i \in \mathbb{R}^3$, $i=1,2,3$ is the triplet parent spheres, $\mathbf{x}_n$ is the unknown position of the center of the new sphere, and $R_i$ denote the corresponding radius.

Once the new sphere is placed in its putative position, a check is run to see if it overlaps with any other spheres. If it does, then the current putative location is discarded, and the sphere is matched with the next set of possible parents. Once the sphere is placed and doesn't intersect with any other spheres, its location and contacts are stored, and the next sphere's radius is randomly drawn from the given distribution. If a given new sphere does not fit with any of the parents, it is discarded, and a new sphere radius is randomly drawn.

We note that in Method 1, unlike the following Method 2, each edge and face of the unit brick is covered with a different random set of spheres.

%We also note that the unit brick can be chosen to coincide with the full domain $\mathcal{V}$ to obtain a non-symmetric random filling.

\subsection{Method 1: filling a parallelepiped-shaped volume} \label{sec:FullMethod1:vol}

When a unit brick  is filled, it may be used ``as is" to represent the full domain $\mathcal{V}$ filled with spheres in a non-symmetric random manner, or as a building block to build larger volumes in a relatively small amount of computing time, by exploiting the symmetry of the unit brick. Method 1 of volume filling is employed to generate a parallelepiped of size $m\times n\times p$, where $m$, $n$ and $p$ are the
numbers of unit bricks in $x$, $y$ and $z$ direction respectively.

In Method 1, information about the spheres that are in contact with the unit brick's faces is used to fill the desired volume by reflecting the unit brick symmetrically with respect to its face planes, so that spheres on the faces would be in direct contact with spheres in a symmetric copy of the unit brick. This process is repeated until desired length is met, and then is repeated in the perpendicular directions. Thus instead of a time-consuming process of space filling with spheres, the coordinates of spheres in additional bricks are computed simply through symmetry transformations of coordinates of spheres in the original unit brick, and radii and contact information are copied directly.

\subsection{Method 1: the main sphere fitting function} \label{sec:FullMethod1:funct}

The \verb|Matlab| function \verb|Method1GenerateSpheres.m| implements Method 1 of unit brick generation and volume filling as described above. The input and output parameters are given in the order of appearance.

\medskip\noindent{\textbf{Input parameters}}:

\begin{itemize}\setlength\itemsep{0ex}
  \item \verb|ProbabilityDistr|: a \verb|Matlab| probability distribution of the radii of spherical particles. This object can be created using the \verb|Matlab| \verb|makedist| function. %[An example: Weibull distribution.]
      For a given distribution, we call $\bar{r}$ the average sphere radius.
  \item \verb|FaceGoal|: fraction of the unit brick face area covered by spheres in contact with it. %[An example: 0.8.]
  \item \verb|BodyGoal|: fraction of the unit brick volume filled by spheres. %[An example: 0.55.]
  \item \verb|SphereContactParameter|: The contact parameter $\varepsilon$, within $[0,1]$. If two spherical particles are within $\verb|SphereContactParameter|\times \bar{r}$ of each other, they are considered to be in contact. %[An example: 0.1.]

  \item \verb|ParentParameter|: the ``parent parameter" $\delta$, within $[0,1]$. If two spherical particles are within
        $\verb|ParentParameter|\times \bar{r}$ of each other, they are considered to be potential ``parents" to further particles.
        %[An example: 0.5.]
  \item \verb|BrickSideLengths|: an array of three values corresponding to absolute lengths (in physical units) of the unit brick sides alonf $x$, $y$, and $z$.
      %[An example: $[30\times \bar{r}, 20\times \bar{r}, 55\times \bar{r}]$.]

  \item \verb|BrickNumbers|: an array of three integer values specifying numbers of copies of the unit brick in $x$, $y$, and $z$ directions required to build the total volume $\mathcal{V}$.
      %  [An example: $[2, 3, 2]$.]
\end{itemize}

The function \verb|Method1GenerateSpheres.m| uses the prescribed probability distribution to fill with spheres the parallelepiped $\mathcal{V}$ having physical dimensions in $x$, $y$ and $z$ directions given by
\beq\label{eq:M1:dom:sizes}
\barr
L_i=\verb|BrickSideLengths(1)|\times \verb|BrickNumbers(i)|,\quad i=1, 2, 3.
\earr
\eeq

\medskip\noindent{\textbf{Output parameters}}:

\begin{itemize}\setlength\itemsep{0ex}
	\item \verb|FinalNSpheres|: The total final number of spheres in the total domain $\mathcal{V}$.
	\item \verb|UnitBrickNSpheres|: The number of spheres in each unit brick.
	\item \verb|Positions|: a matrix
\[\begin{pmatrix}
x_1 & \cdots & x_n & \cdots\\
y_1 & \cdots & y_n & \cdots\\
z_1 & \cdots & z_n & \cdots
\end{pmatrix}
\]
of dimension $3\times \verb|FinalNSpheres|$; the first row stores the $x$-coordinates, second row stores the $y$-coordinates, and the third row the $z$-coordinates of all spheres in $\mathcal{V}$. Thus the $n^{\mathrm{th}}$ column of the \verb|Positions| matrix gives the coordinates of the $n^{\mathrm{th}}$ sphere.
	\item \verb|Radii|: a $1\times \verb|FinalNSpheres|$ matrix that stores the radii of all spheres in the whole domain $\mathcal{V}$. The $n^{\mathrm{th}}$ entry is the radius of the $n^{\mathrm{th}}$ sphere.
	\item \verb|Contacts|: keeps track of which particles are in contact. This matrix consists of two columns; a pair of entries in the same row is the pair of indices of two spheres that are in contact.
	\item \verb|ListXmin|, \verb|ListYmin|, \verb|ListZmin|, \verb|ListXmax|, \verb|ListYmax|, \verb|ListZmax|: single-column matrices, each storing all indices of spherical particles in the \verb|Positions| matrix that are in contact with the respective boundaries of the total domain $\mathcal{V}$ corresponding to minimal $x$, minimal $y$, minimal $z$, maximal $x$, maximal $y$, and maximal $z$.
\end{itemize}

\subsection{Method 1: a typical program sequence and run examples} \label{sec:FullMethod1:eg}

As a run example for the first method of volume filling, we choose the Weibull distribution \cite{SteubenJohnC2016Demo,spierings2011influence} for the sphere radii, given by the probability density function (PDF)
\begin{equation}\label{Weib}
    f(r; \lambda , k) = \frac{k}{\lambda}\left ( \frac{r}{\lambda} \right )^{k-1} e^{-(r/ \lambda)^k},
\end{equation}
where $r\geq 0$ is the dimensional random variable describing the sphere radius, $\lambda >0$ is the scale parameter measured in the same length units as the random variable $r$, and $k>0$ is the dimensionless shape parameter. The distribution \eqref{Weib} has the mean value
\begin{equation}\label{Weib:mean}
\bar{r} = \lambda \;\Gamma\left(1+\dfrac{1}{k}\right),
\end{equation}
where $\Gamma$ is the gamma function. For the current example, we choose random sphere parameters corresponding to powder bed 3D printing process with 316L stainless steel powder \cite{SteubenJohnC2016Demo},
\begin{equation}\label{Weib:params}
\lambda = 15.7~\text{\textmu m},\qquad k = 3.55, \qquad \bar{r}\simeq 14.14~\text{\textmu m}.
\end{equation}
We note that in the literature, in particular, that devoted to additive manufacturing, diameters of spherical particles are often used instead of radii. For example, the diameter-based value $\lambda = 31.4~\text{\textmu m}$ is used in Ref.~\cite{SteubenJohnC2016Demo} (see Ref.~\cite{SteubenJohnC2016Demo} Table 1, steel sample S2).

The \verb|Matlab| script \verb|Example1A_Method1_Generate_and_Plot.m| listed in Appendix \ref{appendix:eg1:generate} below specifies the probability distribution \eqref{Weib}, and defines the main parameters for the run, including the characteristics of the domain to be filled with spheres, and the variables controlling the sphere sizes \eqref{Weib:params} and the surface and volume fill ratios. The description of some commands and the run parameters used in the script are also listed in Appendix \ref{appendix:eg1:generate}. The script calls the main volume filling function \verb|Method1GenerateSpheres.m|, saves the data, and plots the resulting graphs.

\bigskip\noindent \textbf{Example 1A.} In the first example run for Method 1, the following input parameters were used.
\begin{itemize} \setlength\itemsep{0ex}
  \item \verb|ProbabilityDistr|: Weibull \eqref{Weib}, \eqref{Weib:params}.
  \item \verb|FaceGoal|: 0.8.
  \item \verb|BodyGoal|: 0.55.
  \item \verb|SphereContactParameter|: 0.2.
  \item \verb|ParentParameter|: 0.5.
  \item \verb|BrickSideLengths|: \verb|[1; 1; 1]*std_length|, where $\verb|std_length|=15\,\bar{r}$.
  \item \verb|BrickNumbers|: \verb|[2; 2; 1]|.
\end{itemize}

We note that a reference value for the face goal parameter can be computed as the ratio of the sphere projection area to the area of a square with side length $2r$. Similarly, the body goal is estimated as the ratio of sphere to circumscribed brick volume ratio, which yields
\beq\label{eq:sizes:goals}
\barr
\verb|FaceGoal|\sim (\pi r^2)/(2r)^2 = \pi/4 \simeq 0.78,\\
\verb|BodyGoal|\sim (4\pi r^3/3)/(2r)^3\simeq 0.52.
\earr
\eeq
Values of \verb|FaceGoal| and \verb|BodyGoal| optimal for a specific application can be determined experimentally.

\begin{figure}[htbp]
\begin{subfigure}[b]{.5\textwidth}
  \centering
  \subcaptionbox{\label{fig:del_1_var_diff}}{\includegraphics[width=0.85\textwidth]{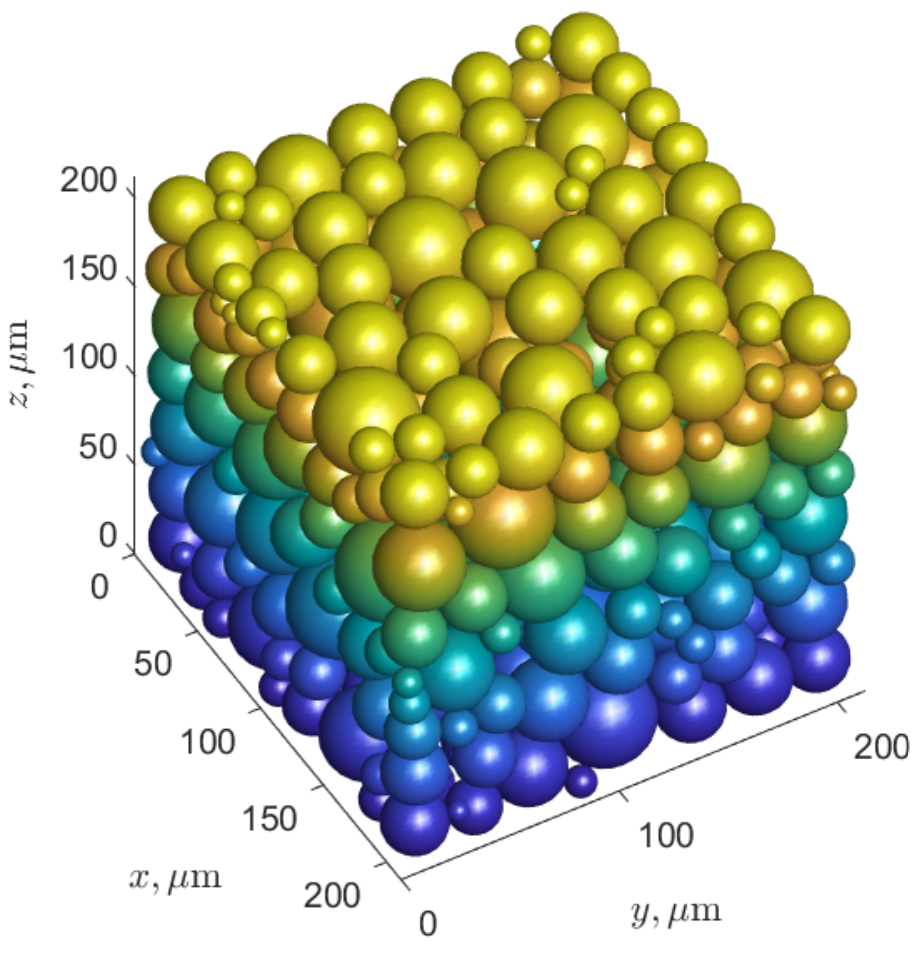}}
\end{subfigure}
\begin{subfigure}[b]{.5\textwidth}
\centering
\subcaptionbox{\label{fig:del_2_var_diff}}{\includegraphics[width=0.85\textwidth]{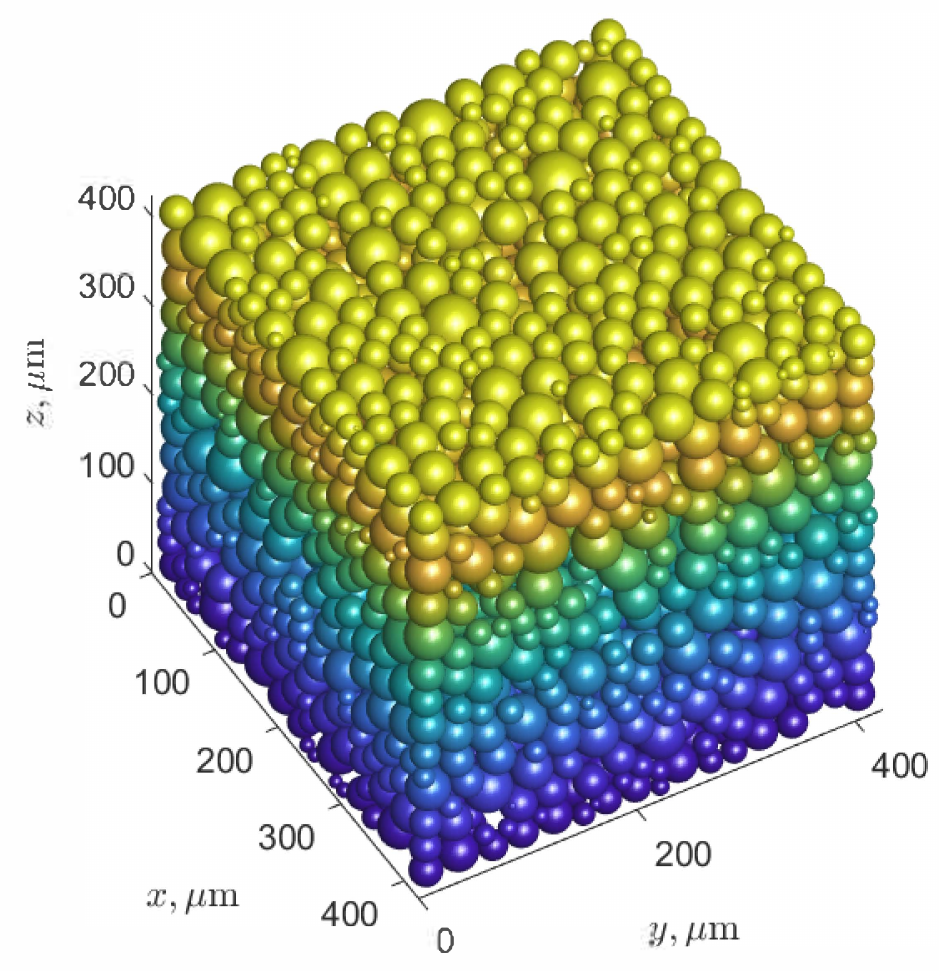}}
\end{subfigure}
\\
\begin{subfigure}[b]{.5\textwidth}
\centering
\subcaptionbox{\label{fig:del_3_var_diff}}{\includegraphics[width=0.85\textwidth]{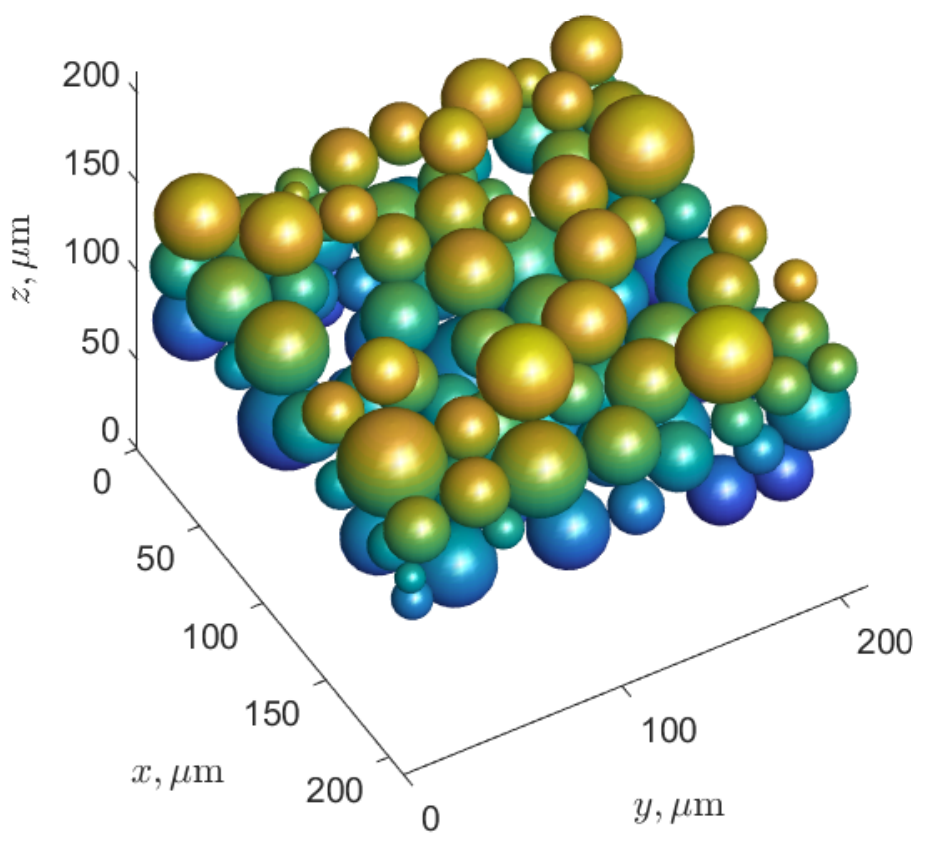}}
\end{subfigure}
\begin{subfigure}[b]{.5\textwidth}
\centering
\subcaptionbox{\label{fig:del_4_var_diff}}{\includegraphics[width=0.85\textwidth]{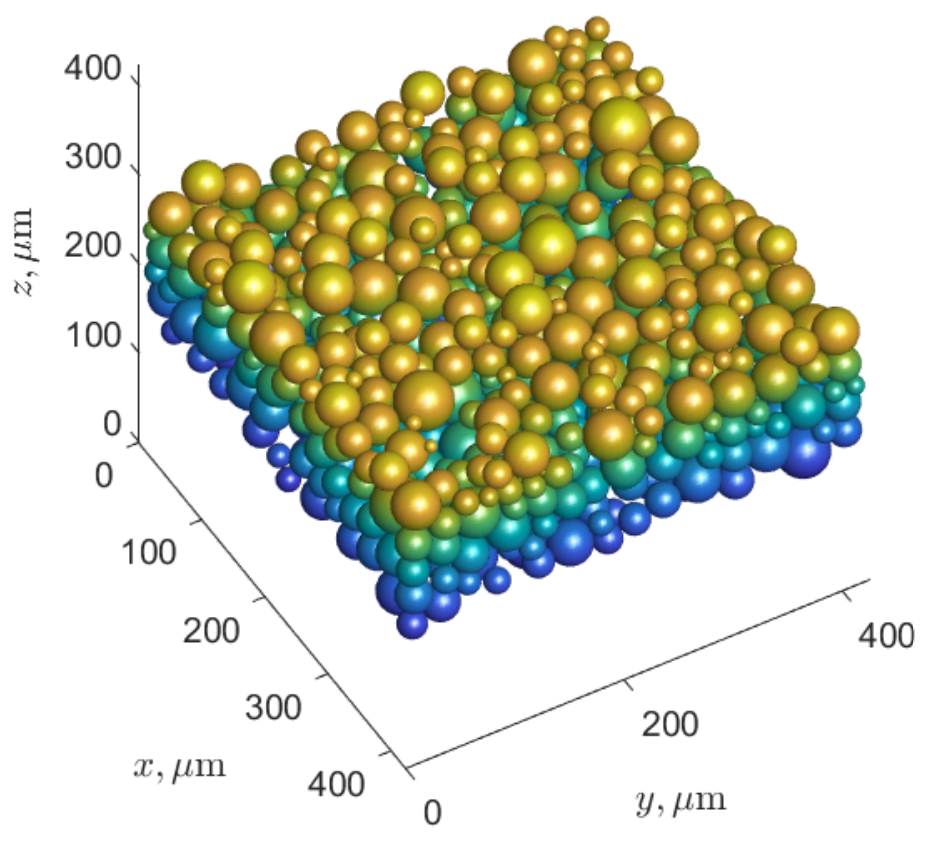}}
\end{subfigure}
\\
\begin{subfigure}[b]{.5\textwidth}
  \centering
  \subcaptionbox{\label{fig:del_5_var_diff}}{\includegraphics[width=0.85\textwidth]{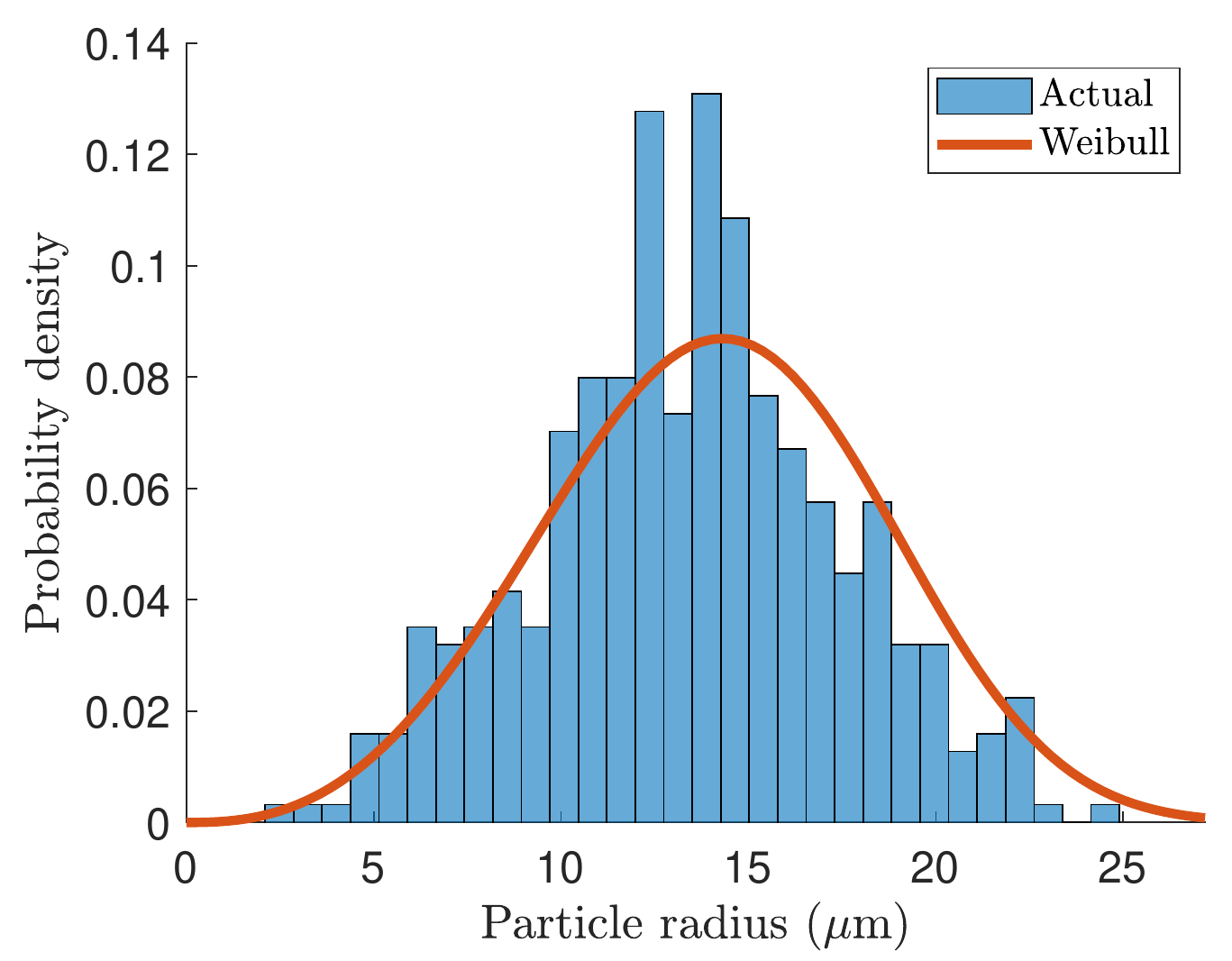}}
\end{subfigure}
\begin{subfigure}[b]{.5\textwidth}
\centering
\subcaptionbox{\label{fig:del_6_var_diff}}{\includegraphics[width=0.85\textwidth]{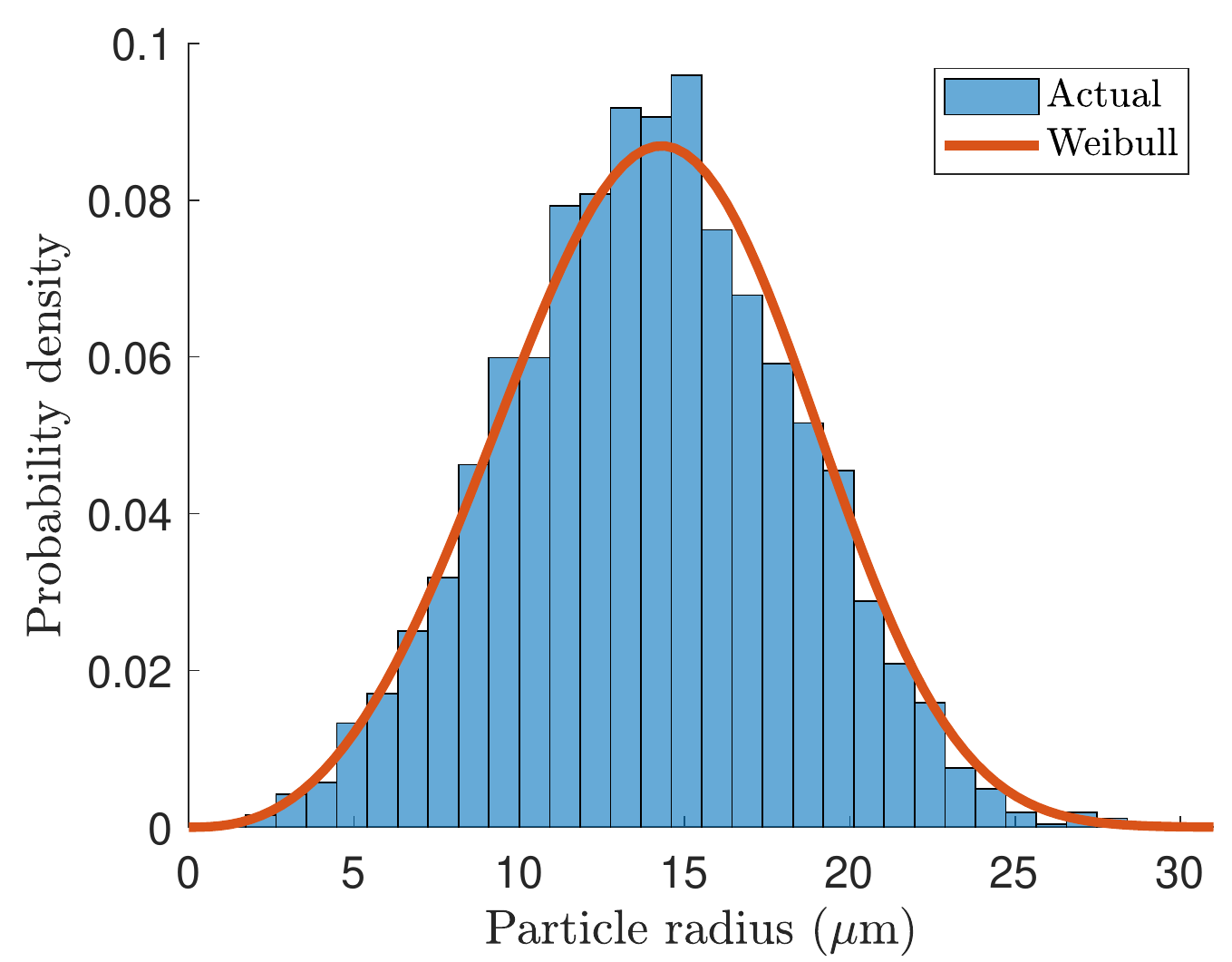}}
\end{subfigure}
\caption{\footnotesize (a) Example 1A: a unit cube with side length $15\,\bar{r}$ filled with spheres using Method 1 with and Weibull distribution \eqref{Weib}, \eqref{Weib:params}. (c) Spheres in middle $1/3$ of the cube, showing the internal structure of the filled cube. (e) Actual sphere radius distribution in the obtained sample, compared to the given Weibull distribution \eqref{Weib}, \eqref{Weib:params}. (b,d,f) Same plots Example 1B: a unit cube with side length $30\,\bar{r}$.
}
\label{fig:Method1:all}
\end{figure}

The sphere generating script  \verb|Example1A_Method1_Generate_Plot.m| is also used to produce the plots and save figure files for the current example. Figure \ref{fig:Method1:all}\,(a, c, e) shows the unit brick, its internal structure, and the histogram of actual particle sizes compared to the probability density of the given distribution \eqref{Weib} for Example 1A.

% Here for both eg A nad B, contact parameter 02 was used.
%re-done in v17 to correct sizes, when d->radius

\bigskip\noindent \textbf{Example 1B.} Here we use the same setup as in Example 1A, with a larger unit brick side length:
\[
\verb|std_length|=30\,\bar{r},
\]
and consequently, four times as many spheres per unit cube. The unit cube and the sphere size histogram for this example are shown in Figure \ref{fig:Method1:all}\,(b, d, f). In particular, the actual sphere size distribution histogram in Figure \ref{fig:Method1:all}\,(f) is closer to  the given Weibull distribution than that for Example 1A (Figure \ref{fig:Method1:all}\,(e)); this is due to an increased freedom of fitting random-sized spheres into a unit cube that is larger (relative to $\bar{r}$) than that in Example 1A.

\begin{figure}[!htbp]
\begin{subfigure}[c]{.5\textwidth}
\centering
\subcaptionbox{\label{fig:del_3_var_diff}}{\includegraphics[width=\textwidth]{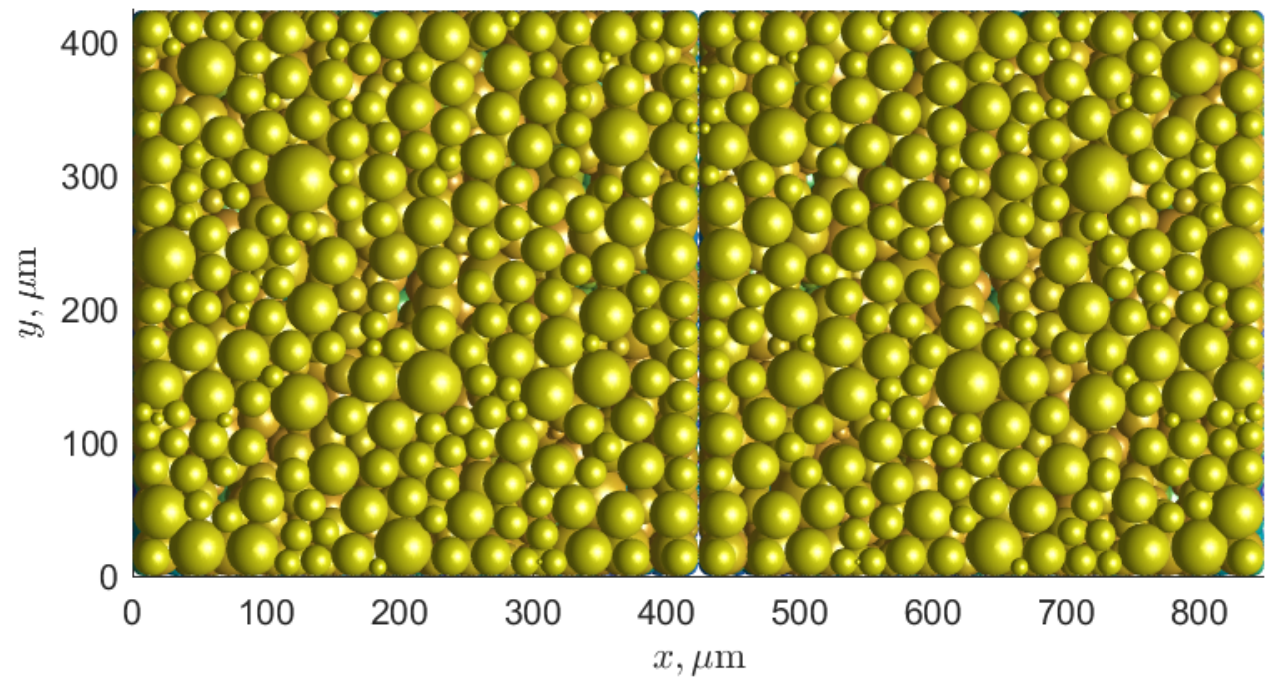}}
\end{subfigure}
\begin{subfigure}[c]{.5\textwidth}
\centering
\subcaptionbox{\label{fig:del_4_var_diff}}{\includegraphics[width=\textwidth]{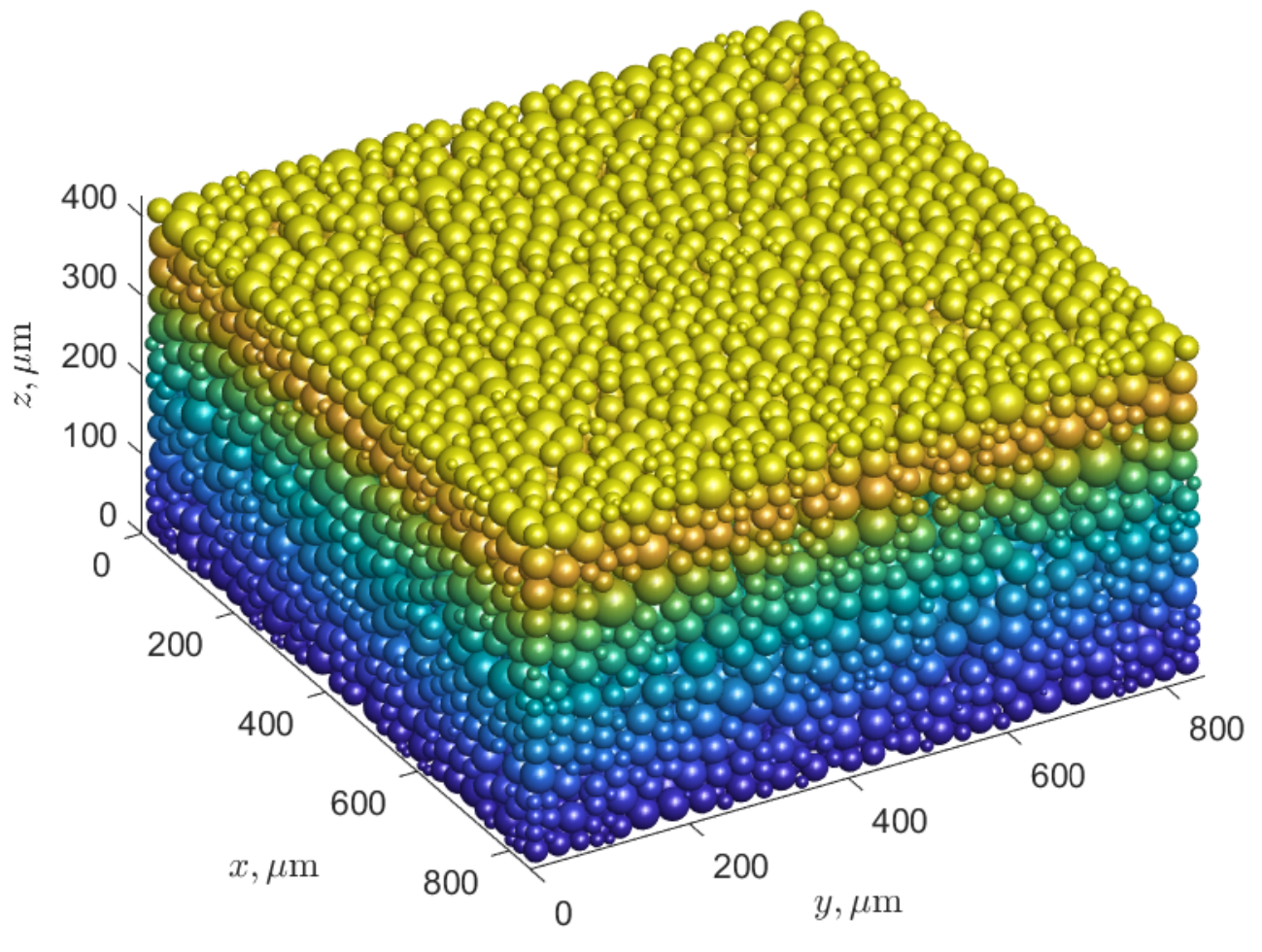}}
\end{subfigure}\\
\begin{subfigure}[c]{.99\textwidth}
  \centering
  \subcaptionbox{\label{fig:del_1_var_diff}}{\includegraphics[width=0.7\textwidth]{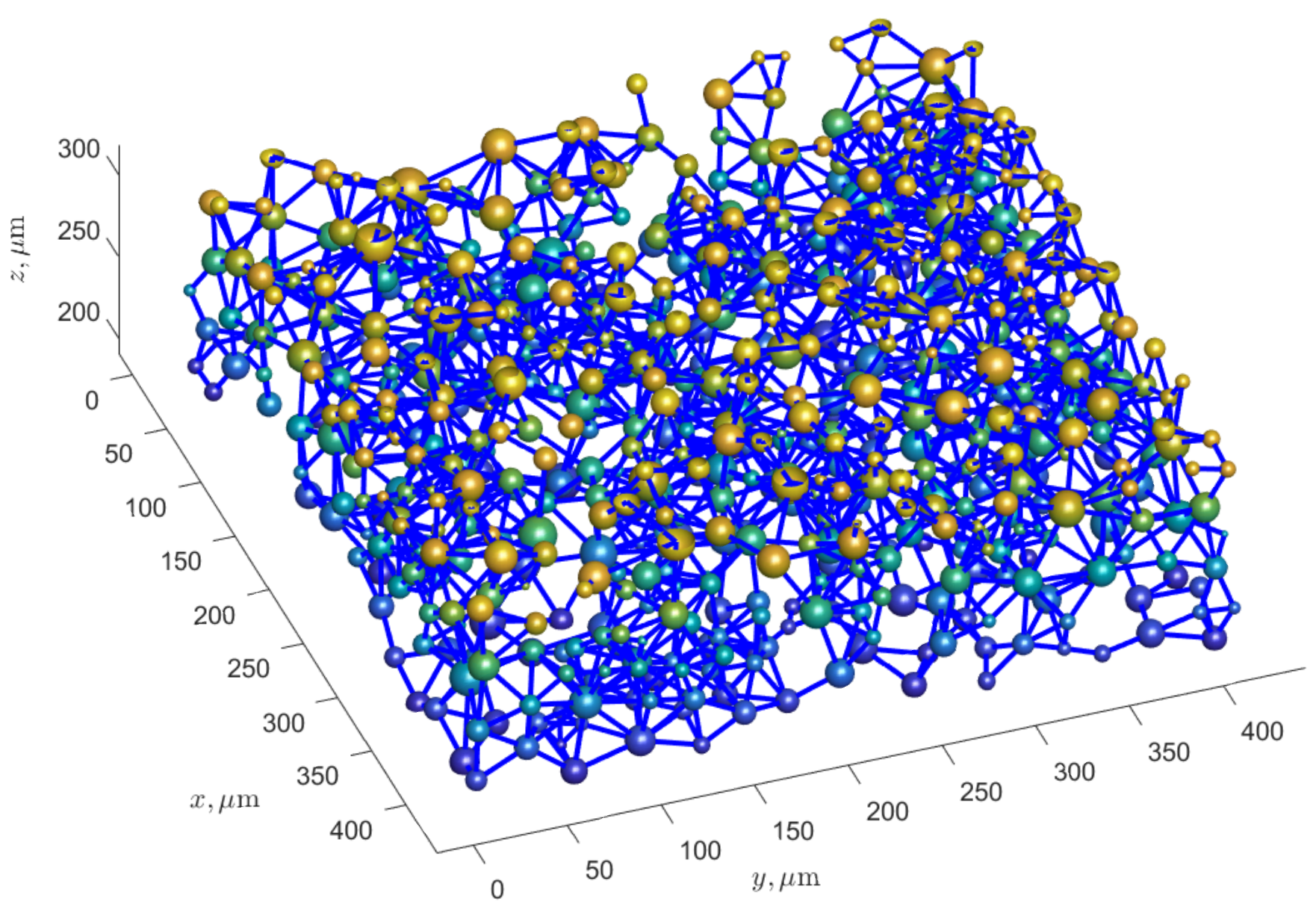}}
\end{subfigure}

\caption{\footnotesize (a) Example 1B: two unit cubes joined to form a $2\times 1$ row - top view. (b) Two rows (four unit cubes) joined to form the final domain $\mathcal{V}$. (c) The connectivity graph for a horizontal slab subdomain of the unit cube of Example 1B (the spheres are scaled down to 40\% of their actual sizes to visualize connections).
}
\label{fig:Method1:2:Full:conn}
\end{figure}

Figure \ref{fig:Method1:2:Full:conn} show the construction of the total volume $\mathcal{V}$ made of $2\times 2\times 1$ unit cubes, and the connectivity graph joining pairs of particles that are in contact, located within the horizontal slab $z = (0.6 \pm 0.3) \times \verb|std_length|$.

%re-done in v17 to correct sizes, when d->radius. Remains: insert new Connectivity graph

\medskip\noindent{\textbf{Examples 1A and 1B: computation times}}. Computations were performed on Matlab 2021a, using a Dell workstation with two Xeon processors, 32 logical processors, and 128 GB memory.  The computation times listed in Table \ref{table:EG:Mehtod1:times} below depend on the dimensions size of the unit brick and the numbers of unit bricks \verb|BrickNumbers| along the axes to form the total domain $\mathcal{V}$. A strong dependence of the computation times on \verb|SphereContactParameter| $\varepsilon$ is observed. (All computations were also tested on an Intel i7-based laptop with one physical and four logical processors and 16 GB memory, resulting on average in 1.5 to three times longer computations).

%==========================================
% on Lomonosov,  $T = 74$ min -> 21 min... 3+ times faster
\begin{table}[htbp]
\begin{center}
\begin{tabular}{|c|c|c|c|}
\hline \vspacebefore
\multirow{2}{*}{Cube side length} & \multicolumn{2}{c|}{Sphere contact parameter} \\[0.5ex]
&0.1 &0.2\\
\hline\hline \vspacebefore

$15\, \bar{r}$ (Example 1A) & $T = 6$~\text{min},~ $N=450$  & $T = 2$~\text{min},~ $N=412$  \\% Laptop  6 min, 3 min
$30\, \bar{r}$ (Example 1B) & $T = 187$~\text{min},~ $N=3075$ & $T = 40$~\text{min},~ $N=2867$\\%  Laptop  ** min, 40 min

\hline
\end{tabular}
\end{center}
\caption{\footnotesize Sample desktop computation times $T$ numbers of spheres $N$ in a unit cube for Examples 1A and 1B, for the total domain $\mathcal{V}$ built of $2\times 2 \times 1$ unit cubes of different side lengths, for two different values of the sphere contact parameter.
}
\label{table:EG:Mehtod1:times}
\end{table}
%==========================================
We also note that plotting can take a relatively long time, similar to the computation time of sphere filling, due to a large number of spherical particles, each represented by a graphical object with multiple faces  in the \verb|Matlab| \verb|sphere| plotting routine.

\bigskip\noindent \textbf{Example 1C: non-cubical unit bricks.} In the current example, we use a function similar to Example 1's \verb|Example1A_SpherePackingMethod1_Generate.m| to call the main sphere fitting function \verb|Method1GenerateSpheres.m| with different parameters:
\begin{itemize}\setlength\itemsep{0ex}
  \item \verb|ProbabilityDistr|: Weibull distribution  \eqref{Weib}, \eqref{Weib:params} of sphere radii,
  \item \verb|FaceGoal|: 0.8,
  \item \verb|BodyGoal|: 0.55,
  \item \verb|SphereContactParameter|: 0.2,
  \item \verb|ParentParameter|: 0.5,
  \item \verb|BrickSideLengths|: \verb|[1.2; 1.7; 1]*std_length|, where $\verb|std_length|=15\,\bar{r}$,
  \item \verb|BrickNumbers|: \verb|[4; 4; 2]|.
\end{itemize}
As a result, the total domain $\mathcal{V}$ of size $4\times 4\times 2\,=\,32$ unit bricks is filled with spheres. Each unit brick is non-cubical, with size lengths specified in \verb|BrickSideLengths|.

\begin{figure}[!htbp]
\begin{subfigure}[b]{.5\textwidth}
  \centering
  \subcaptionbox{\label{fig:del_1_var_diff}}{\includegraphics[width=0.95\textwidth]{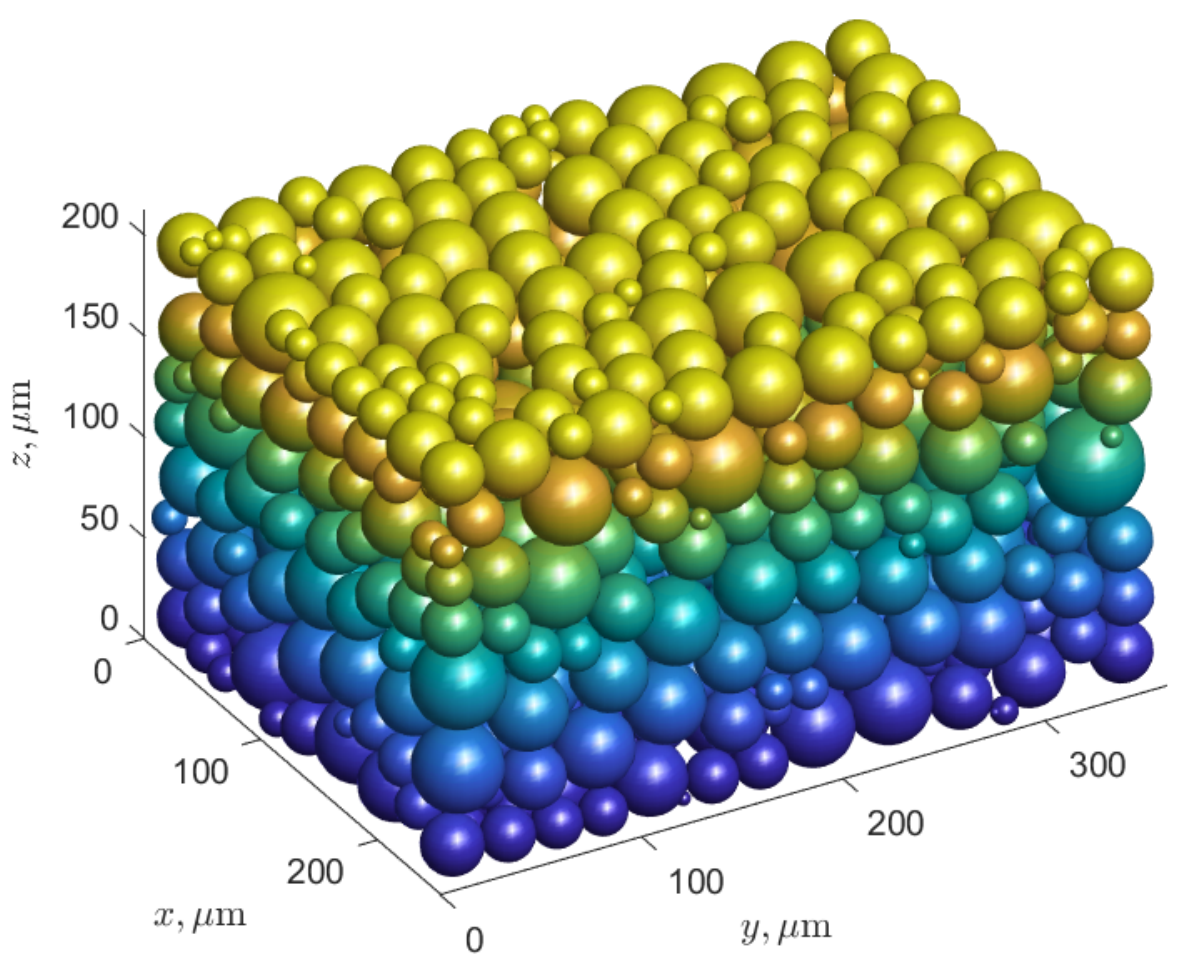}}
\end{subfigure}
\begin{subfigure}[b]{.5\textwidth}
\centering
\subcaptionbox{\label{fig:del_2_var_diff}}{\includegraphics[width=0.95\textwidth]{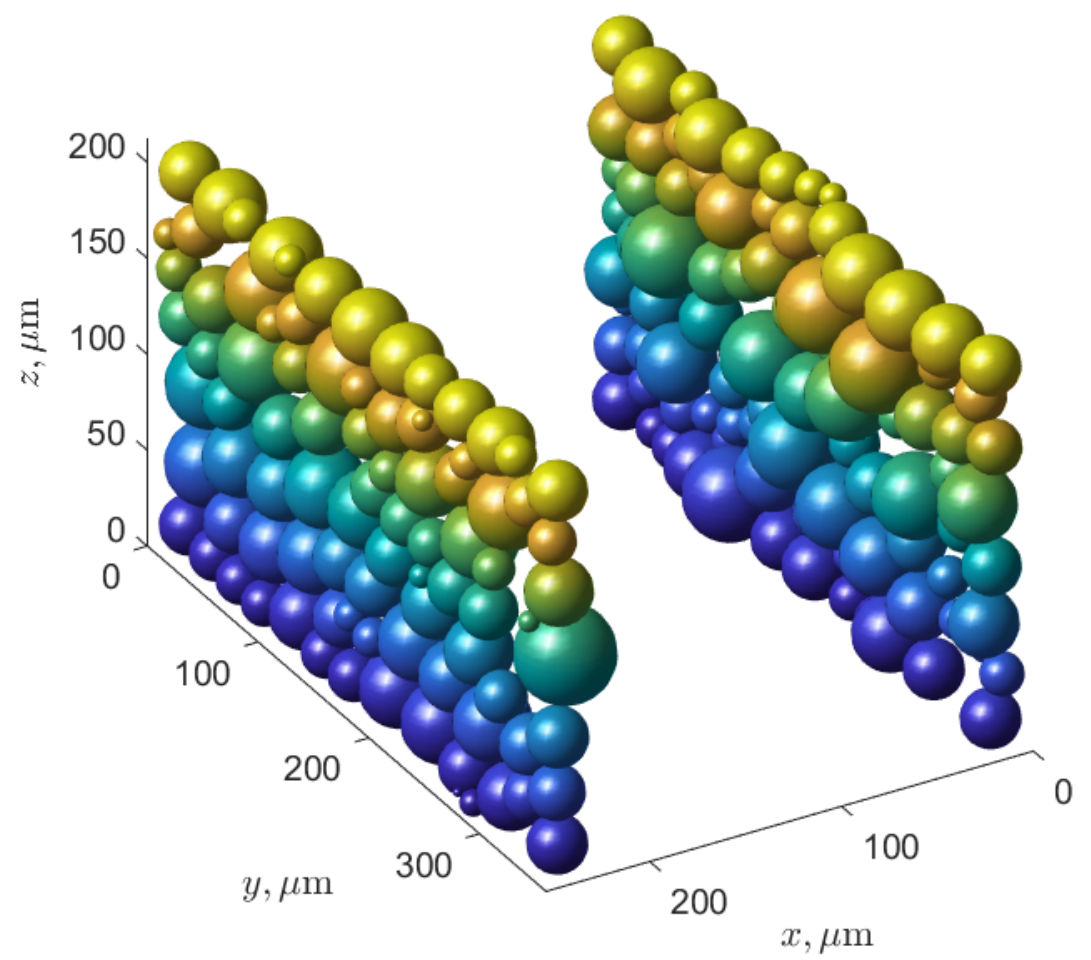}}
\end{subfigure}
\\
\begin{subfigure}[b]{.99\textwidth}
  \centering
  \subcaptionbox{\label{fig:del_1_var_diff}}{\includegraphics[width=0.65\textwidth]{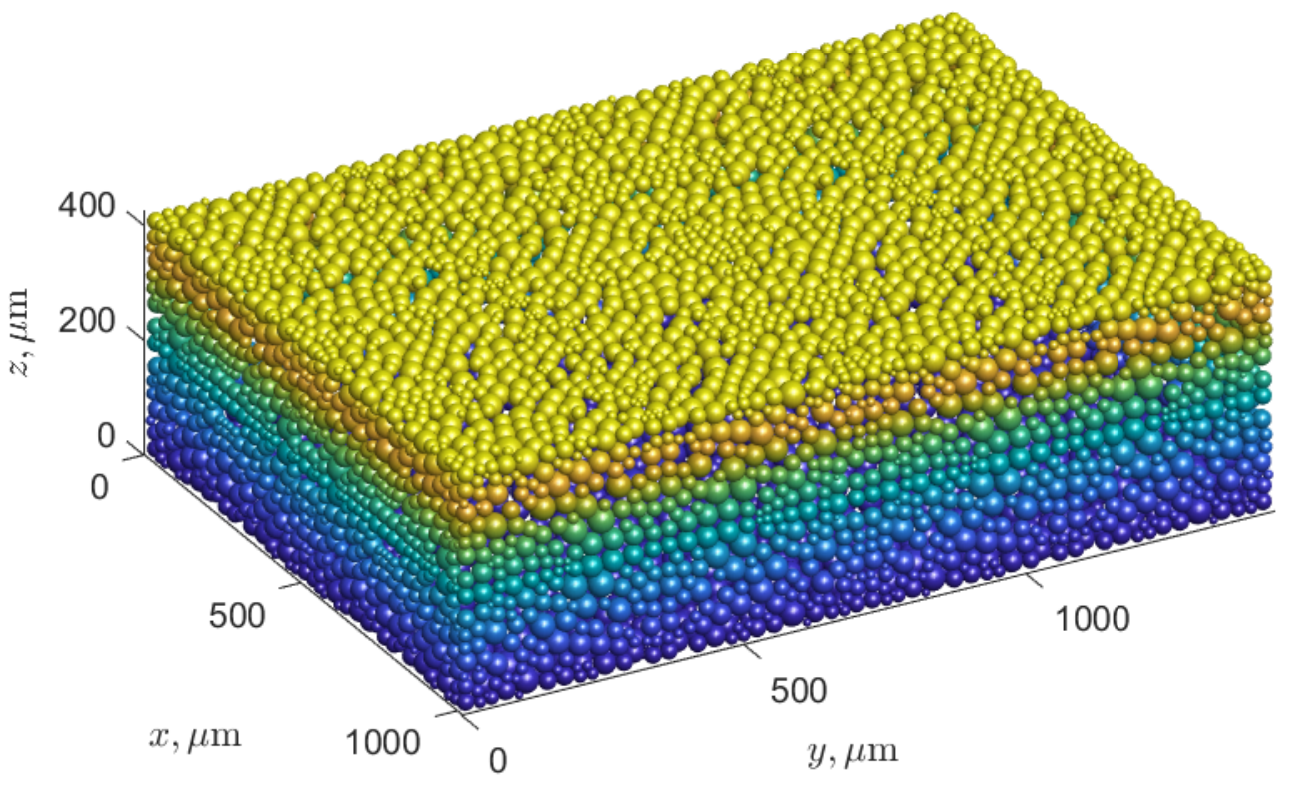}}
\end{subfigure}

    \caption{\footnotesize (a) Example 1C: the unit brick with side lengths $15\,\bar{r}\times [1.2, 1.7, 1]$ filled with spheres using Method 1 with and Weibull distribution \eqref{Weib}, \eqref{Weib:params}. (b) Spheres tangent to the brick sides corresponding to the minimal and the maximal $x$. (c) the total domain $\mathcal{V}$: $4\times 4\times 2$ unit bricks.
}
\label{fig:Method1:Eg1C}
\end{figure}

Figure \ref{fig:Method1:Eg1C} shows the resulting unit brick and the spheres touching the sides corresponding to the minimal and the maximal $x$-value, as well as the total build of the domain $\mathcal{V}$.

The total computation of the Method 1 sphere filling of the total domain $\mathcal{V}$ using 32 unit bricks specified above, on the same hardware/software configuration took about 10 minutes. %on Lomonosov
For comparison, if the same volume $\mathcal{V}$ is constructed from four larger bricks instead, that is,
\begin{itemize}\setlength\itemsep{0ex}
  \item \verb|BrickSideLengths|: \verb|[1.2; 1.7; 1]*std_length|, where $\verb|std_length|=30\,\bar{r}$,
  \item \verb|BrickNumbers|: \verb|[2; 2; 1]|,
\end{itemize}
the computation time is increased to approximately 160 minutes. %Lomonosov - 9.6827e+03 sec

%*** Lomonosov, file Example1C_SpherePackingMethod1_Generate_221_30.m
% results: Example1C_SpherePackingMethod1_Results221_30.mat

\newpage

%==================
\section{Filling a domain with spheres: Method 2} \label{sec:FullMethod2}

Similarly to the first method, the second method is used to fill a parallelepiped-shaped domain $\mathcal{V}$ with one or more unit bricks packed with random spheres whose radii follow a given probability distribution.

One of the main differences in the second method from the first one is the fact that instead of being fully inside of the unit brick and touching the brick faces (Method 1), in Method 2, centers of the spheres on each face are located on the faces themselves. This results in somewhat different-looking unit bricks. Another difference lies in the symmetry of edges and faces. When filling the unit brick, Method 2 starts by placing equal spheres into all corners; then, unlike the first method, only one edge is then filled in each ($x$, $y$, and $z$) direction, and four parallel copies of each edge are made, to fill all 12 edges of the unit brick. Similarly, only three faces (in $xy$, $yz$, and $xz$ planes) are filled with random spheres, and these facies are copied onto the opposite ones. Finally, the unit brick volume is filled with random spheres. As a result of this procedure, a unit brick with symmetric faces but a non-symmetric volume filling is obtained. In order to construct a sphere filling for the total domain $\mathcal{V}$ when it is made of several unit bricks, the unit brick is directly copied as many times as required in $x$, $y$, and $z$ directions; the face symmetry provides a seamless fit when unit bricks are joined together.

\subsection{Method 2: the main sphere fitting function}

Parallel to Method 1, Method 2 is implemented in \verb|Matlab|, in the function named \\
\verb|Method2GenerateSpheres.m|,
which has the same parameters as the Method 1 function (see Section \ref{sec:FullMethod1:funct}). The function uses any prescribed probability distribution, fills the unit brick, and then the domain constructed from unit bricks, as specified. All input and  output parameters in the Method 2 sphere fitting function also coincide with those for Method 1, making the two methods fully interchangeable, yet leading to different fillings.

\subsection{Method 2: a typical program sequence and a run example} \label{sec:FullMethod2:eg}

\noindent \textbf{Example 2.} In the current example for Method 2, we chose a different
probability distribution for sphere sizes: the Gamma distribution given by the probability density function (PDF)
\begin{equation}\label{Gamm}
    f(r; k, \theta) = \dfrac{1}{\Gamma(k)} \, \dfrac{x^{k-1}}{\theta^k}\, e^{-x/\theta},
\end{equation}
where $r\geq 0$ is the dimensional random variable describing the sphere radius, $\theta>0$ is the scale parameter measured in the same units as the random variable $r$, and $k>0$ is the dimensionless shape parameter. The distribution \eqref{Gamm} has the mean value
\begin{equation}\label{Gamm:mean}
\bar{r} = k\,\theta.
\end{equation}
For the current example, we choose random sphere parameters similar to Example 1 above, so that the average radius approximately matches that of \eqref{Weib:params}
\begin{equation}\label{Gamm:params}
    \theta=7~\text{\textmu m},\qquad k = 2, \qquad \bar{r} = 14~\text{\textmu m}.
\end{equation}

The script \verb|Example2_Method2_Generate_and_Plot.m| (Appendix \ref{appendix:eg2:generate}) was used to call the sphere filling function \verb|Method2GenerateSpheres.m| with the input parameters listed below.
\begin{itemize}\setlength\itemsep{0ex}
  \item \verb|ProbabilityDistr|: Gamma \eqref{Gamm}, \eqref{Gamm:params}.
  \item \verb|FaceGoal|: 1.0.
  \item \verb|BodyGoal|: 0.9.
  \item \verb|SphereContactParameter|: 0.1.
  \item \verb|ParentParameter|: 0.5.
  \item \verb|BrickSideLengths|: \verb|[1; 1; 1]*std_length|, where $\verb|std_length|=30\,\bar{r}$ (see \eqref{Gamm:params}).
  \item \verb|BrickNumbers|: \verb|[2; 2; 1]|.
\end{itemize}
Here the total domain $\mathcal{V}$ is constructed of $2\times 2\times 1$ unit bricks (which are cubes in this example). We note the higher values for \verb|FaceGoal| and \verb|BodyGoal| than used in Method 1 examples. This is natural because for Method 2, centers of boundary spheres are located on the faces, which results in higher relative area and volume occupied by the spheres in the unit brick. Optimal values for face and body goals can be chosen by the user experimentally, depending on a particular application.

\begin{figure}[htbp]
\begin{subfigure}[b]{.5\textwidth}
  \centering
  \subcaptionbox{\label{fig:del_1_var_diff}}{\includegraphics[width=0.85\textwidth]{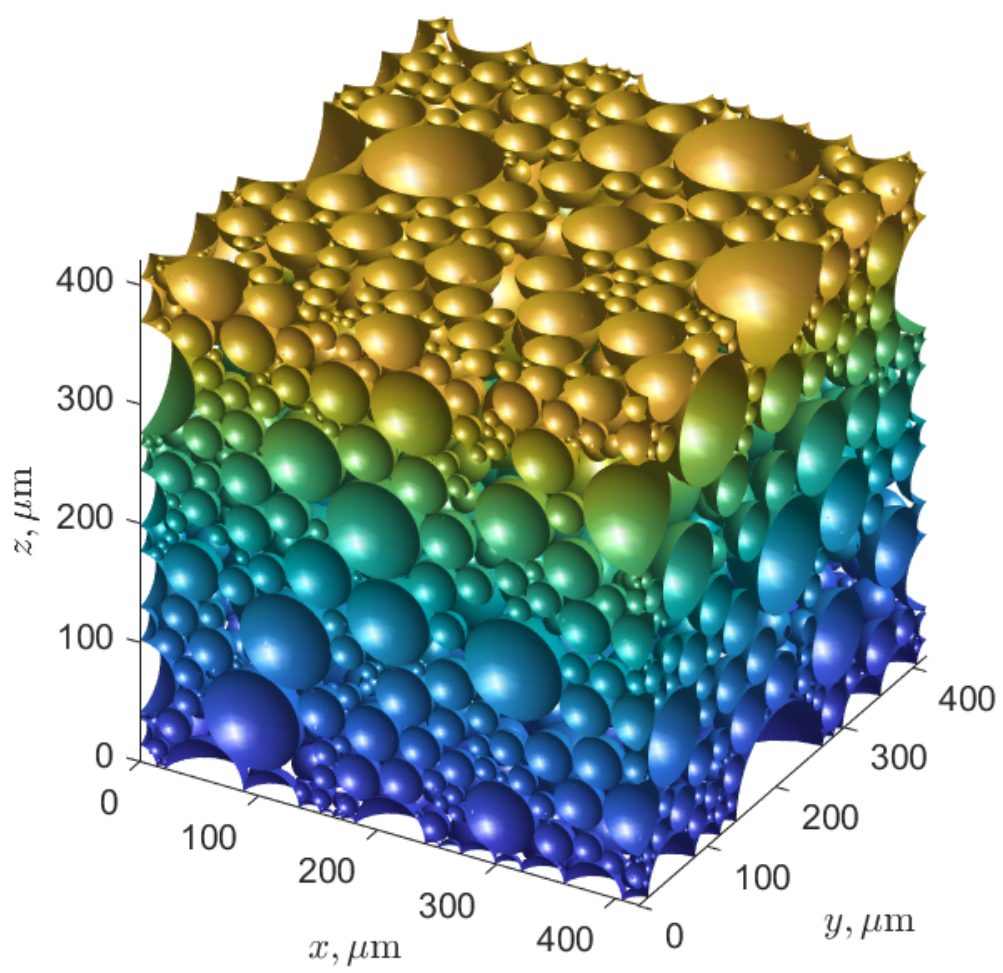}}
\end{subfigure}
\begin{subfigure}[b]{.5\textwidth}
\centering
\subcaptionbox{\label{fig:del_2_var_diff}}{\includegraphics[width=0.85\textwidth]{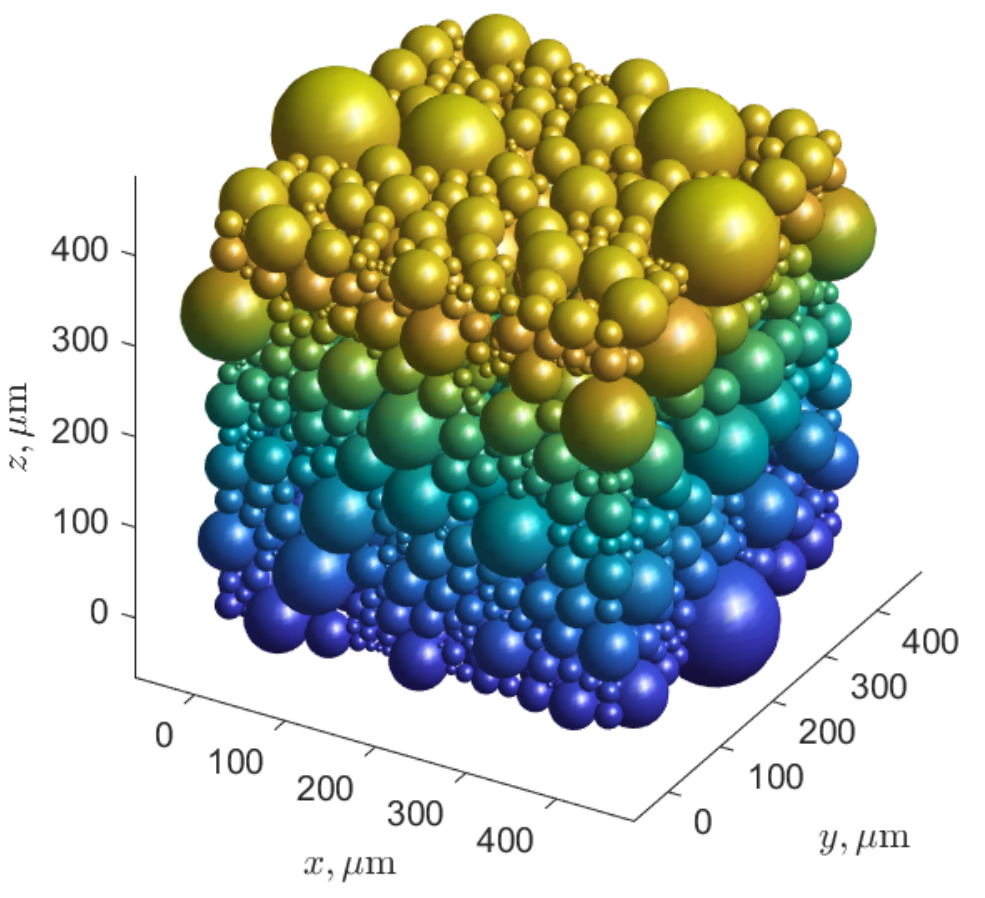}}
\end{subfigure}
\\
\begin{subfigure}[c]{.45\textwidth}
  \centering
  \subcaptionbox{\label{fig:del_3_var_diff}}{\includegraphics[width=0.99\textwidth]{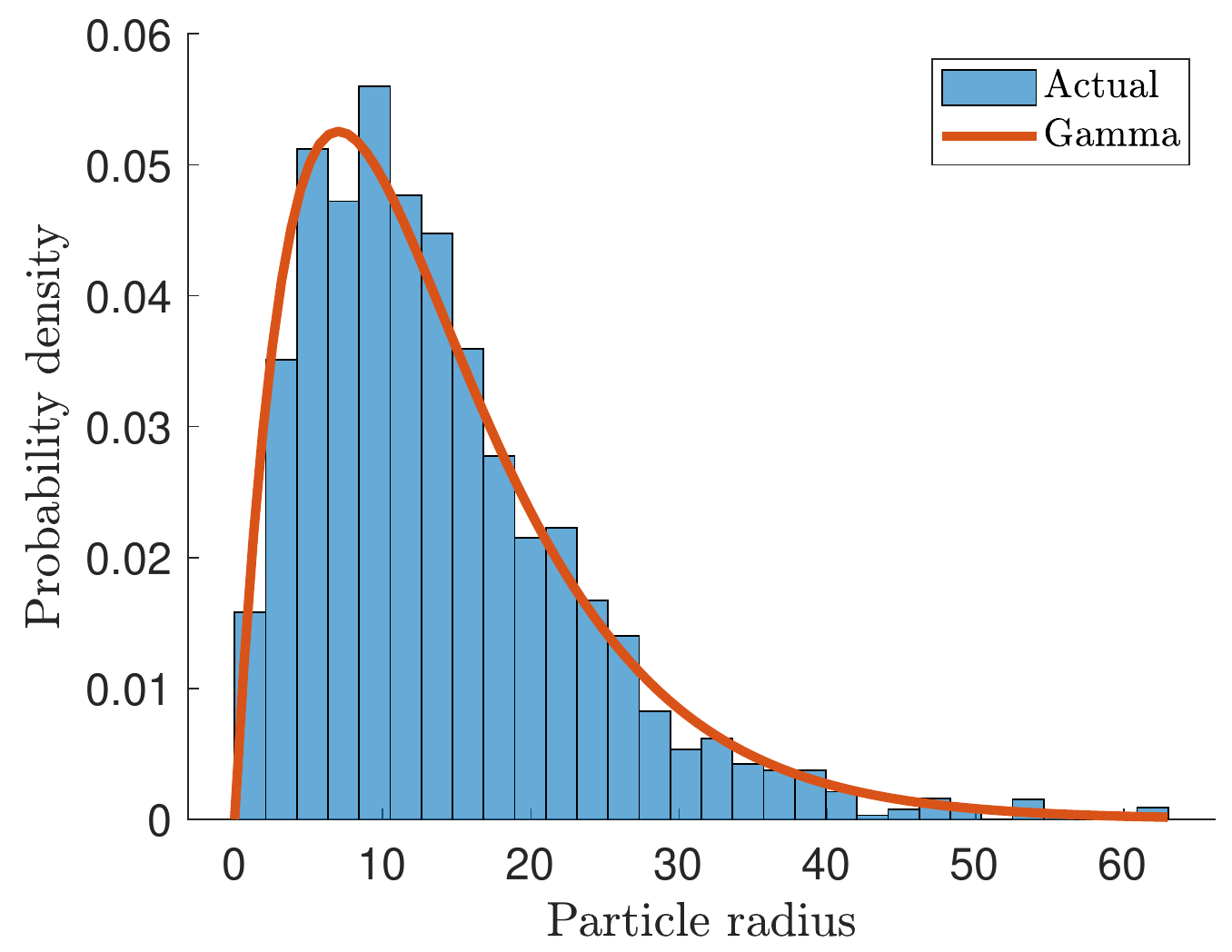}}
\end{subfigure}
\begin{subfigure}[c]{.55\textwidth}
\centering
\subcaptionbox{\label{fig:del_4_var_diff}}{\includegraphics[width=0.99\textwidth]{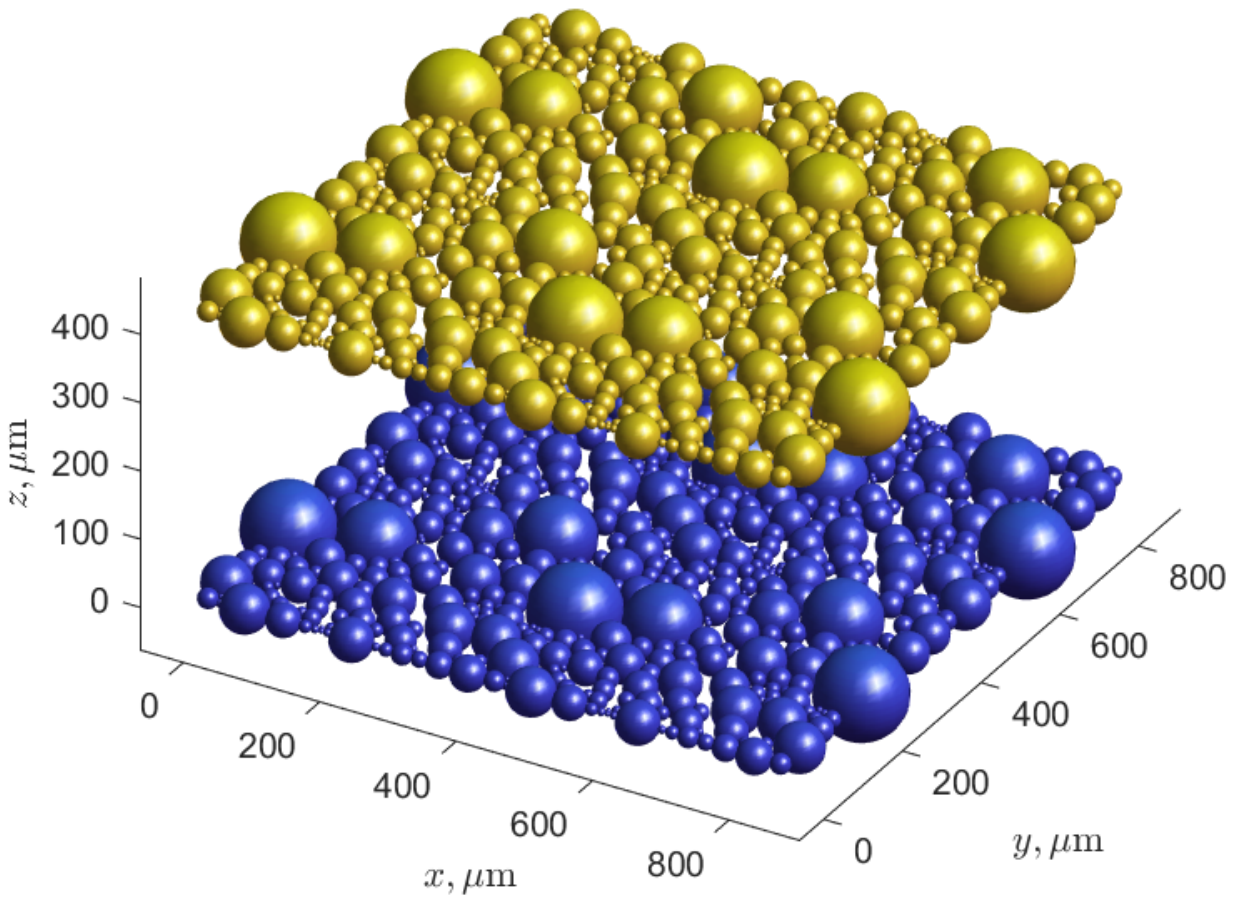}}
\end{subfigure}
\caption{\footnotesize (a) Example 2: a unit brick (cube) with side length $30\,\bar{r}$ filled with spheres using Method 2 with Gamma distribution \eqref{Gamm}, \eqref{Gamm:params} (unit cube clipped to its size, showing the centers of the spheres on the faces). (b) Same as (a) where the full unit cube is shown in a larger domain. (c) The probability function of the theoretical Gamma distribution \eqref{Gamm}, \eqref{Gamm:params} vs.~the histogram of actual particle sizes in the unit cube. (d) The top and bottom of the total volume $\mathcal{V}$ made of $2\times 2\times 1$ unit bricks contain eight identical copies of a horizontal face of the unit cube.
}
\label{fig:Method2:all}
\end{figure}

Figure \ref{fig:Method2:all} shows the unit cube with spheres centered on faces (compare with Figure \ref{fig:Method1:all}\,(a)) and the $z$-boundaries of the total volume $\mathcal{V}$, consisting of the same repeating copies of the unit cube face in the $x=y=0$ plane.

%save('Example2A_SpherePackingMethod1_Results_V221_cubeside15_eps02.mat')
%save('Example2A_SpherePackingMethod1_Results_V221_cubeside15_eps01.mat')
%Example2A_SpherePackingMethod1_Results_V221_cubeside30_eps02.mat
%Example2A_SpherePackingMethod1_Results_V221_cubeside30_eps01.mat

Table \ref{table:EG:Mehtod2:times} contains run times for computations analogous to those performed for Method 1 and listed in Table \ref{table:EG:Mehtod1:times}. Interestingly, for Method 2, unlike Method 1 (cf.~Table \ref{table:EG:Mehtod1:times}), the computation times for the smaller sphere contact parameter $0.1$ are not significantly different, (in fact, are smaller than) the computation times for the contact parameter value $0.2$ (which remains true when the Weibull distribution of Example 1 is used in Method 2 instead og the Gamma distribution).

%========================================== all checked on Lomonosov
\begin{table}[htbp]
\begin{center}
\begin{tabular}{|c|c|c|c|}
\hline \vspacebefore
\multirow{2}{*}{Cube side length} & \multicolumn{2}{c|}{Sphere contact parameter} \\[0.5ex]
&0.1 &0.2\\
\hline\hline \vspacebefore

$15\, \bar{r}$ & $T = 7$~\text{s},~ $N=160$ & $T = 11$~\text{s},~ $N=178$  \\ % laptop 13 s;  ???
$30\, \bar{r}$ & $T = 14$~\text{min},~ $N=1757$ & $T = 15$~\text{min},~ $N=1771$ \\ % laptop 26 min;  Y min
\hline
\end{tabular}
\end{center}
\caption{\footnotesize Sample desktop computation times $T$ numbers of spheres $N$ in a unit cube for Example 2 with Gamma distribution \eqref{Gamm}, \eqref{Gamm:params} , for the total domain $\mathcal{V}$ built of $2\times 2 \times 1$ unit bricks of different side lengths, for two different values of the sphere contact parameter.
}
\label{table:EG:Mehtod2:times}
\end{table}
%==========================================

We have also performed runs of Method 2 with the same Weibull distribution \eqref{Weib}, \eqref{Weib:params} as in Method 1 (resulting graphs are not shown). In this case, using Method 2, the computation times finish faster than those done with Method 1, and the computations also result in bigger numbers of spheres in the unit cube (compare Table \ref{table:EG:Mehtod1:times} for Method 1 with Table \ref{table:EG:Mehtod2:Weib:times} for Method 2).

%========================================== all checked on Lomonosov
\begin{table}[htbp]
\begin{center}
\begin{tabular}{|c|c|c|c|}
\hline \vspacebefore
\multirow{2}{*}{Cube side length} & \multicolumn{2}{c|}{Sphere contact parameter} \\[0.5ex]
&0.1 &0.2\\
\hline\hline \vspacebefore

$15\, \bar{r}$ & $T = 4$~\text{min},~ $N=591$ & $T = 4$~\text{min},~ $N=594$  \\ % laptop 7 min;  6 min
$30\, \bar{r}$ & $T = 32$~\text{min},~ $N=3332$ & $T = 38$~\text{min},~ $N=3606$ \\ % laptop X min;  Y min
\hline
\end{tabular}
\end{center}
\caption{\footnotesize Sample desktop computation times $T$ numbers of spheres $N$ in a unit cube for Example 2 with  Weibull distribution \eqref{Weib}, \eqref{Weib:params}, for the total domain $\mathcal{V}$ built of $2\times 2 \times 1$ unit cubes of different side lengths, for two different values of the sphere contact parameter (cf.~Table \ref{table:EG:Mehtod1:times} for Method 1).
}
\label{table:EG:Mehtod2:Weib:times}
\end{table}
%==========================================

%-------------------------------------

\section{Example 3: a more complex-shaped domain}\label{sec:eg3:hemisph}

The next example builds on the same framework as Methods 1 and 2 of sphere filling, extending to a non-parallelepiped-shaped domain. In this example, the domain $\mathcal{V}$ is similar to a single unit brick in Methods 1 and 2, with the subtraction of two hemispheres centered in the middles of two opposite faces corresponding to $x=0$ and the maximal $x$. The radii of these spheres must be less or equal to $1/2$ of the smallest of the brick dimensions.

A special sphere generating function \verb|Example3GenerateSpheres.m| has been created for this example. The input and output parameters for this function are outlined below. The parameter set for \verb|Example3GenerateSpheres.m| is smaller than that for the sphere generating functions in Methods 1 and 2, but the meaning remains the same (see Section \ref{sec:FullMethod1:funct}).

\medskip\noindent{\textbf{Input parameters}}:

%function [FinalNSpheres, UnitBrickNSpheres, Positions, Radii, Contacts, ListXmin, ListYmin, ListZmin, ListXmax, ListYmax, ListZmax]
%= Method1GenerateSpheres(ProbabilityDistr, FaceGoal, BodyGoal, SphereContactParameter, ParentParameter, BrickSideLengths, BrickNumbers)

\begin{itemize}\setlength\itemsep{0ex}
  \item \verb|ProbabilityDistr|: a \verb|Matlab| probability distribution of the radii of spherical particles.
      The average sphere radius is denoted by $\bar{r}$.
  \item \verb|BrickSideLengths|: an array of three values corresponding to absolute lengths (in physical units) of the unit brick sides along $x$, $y$, and $z$.
  \item \verb|HemisphereRadii|: an array containing two values corresponding to the radii (in physical units) of the two hemispheres centered in the middles of two opposite faces corresponding to $x=0$ and the maximal $x$, defining the computation domain.

  \item \verb|FaceGoal|: fraction of the domain boundary face area covered by projections of spheres in contact with it.
  \item \verb|BodyGoal|: fraction of the domain volume filled by spheres.
  \item \verb|SphereContactParameter|: the contact parameter, within $[0,1]$. If two spherical particles are within $\verb|SphereContactParameter|\times \bar{r}$ of each other, they are considered to be in contact.
\end{itemize}

\medskip\noindent{\textbf{Output parameters}}:

\begin{itemize}\setlength\itemsep{0ex}
	\item \verb|NSpheres|: The total number of random spheres placed into the domain $\mathcal{V}$.
	\item \verb|Positions|: a matrix of coordinates of the random spheres (see Section \ref{sec:FullMethod1:funct}).
	\item \verb|Radii|: a vector storing the radii of all random spheres in the total domain $\mathcal{V}$ (see Section \ref{sec:FullMethod1:funct}).
	\item \verb|Contacts|: A matrix containing sphere pairs that are in contact (see Section \ref{sec:FullMethod1:funct}).
\end{itemize}

\begin{figure}[htbp]
\begin{subfigure}[b]{.5\textwidth}
  \centering
  \subcaptionbox{\label{fig:Eg3:all:A}}{\includegraphics[width=0.95\textwidth]{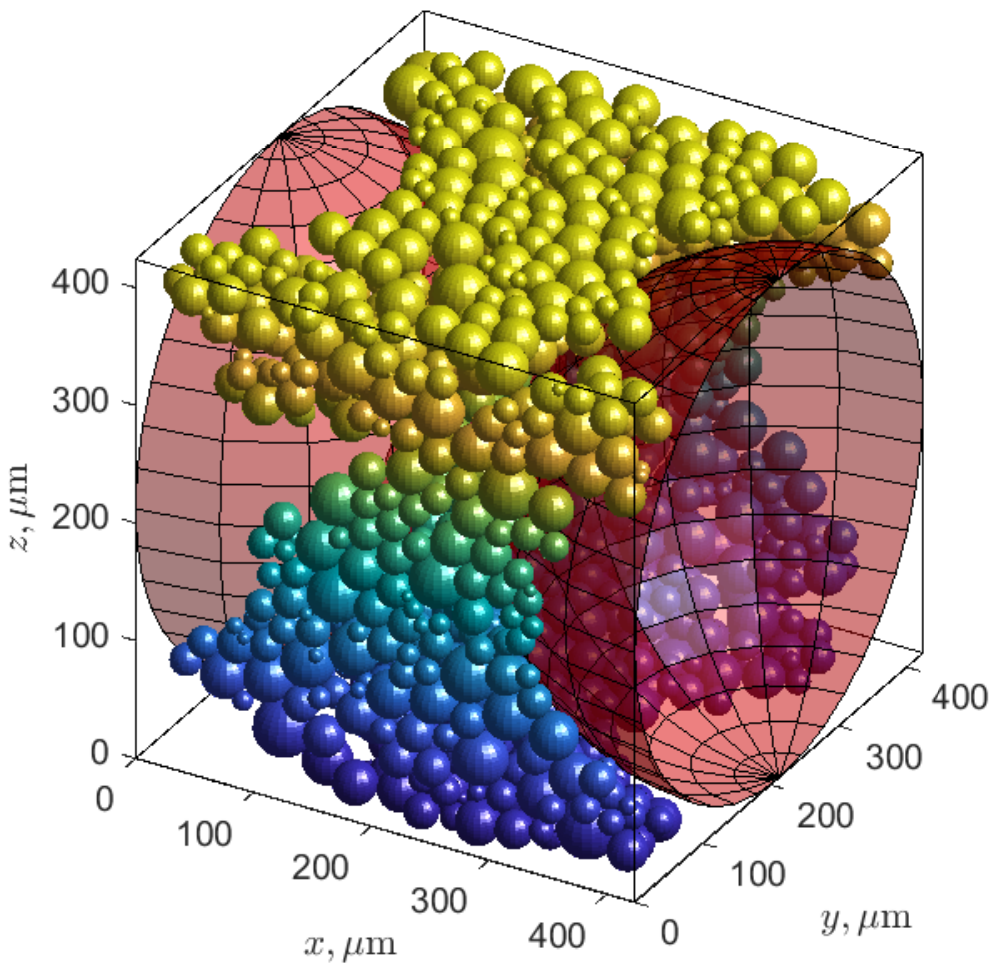}}
\end{subfigure}
\begin{subfigure}[b]{.5\textwidth}
  \centering
  \subcaptionbox{\label{fig:Eg3:all:B}}{\includegraphics[width=0.95\textwidth]{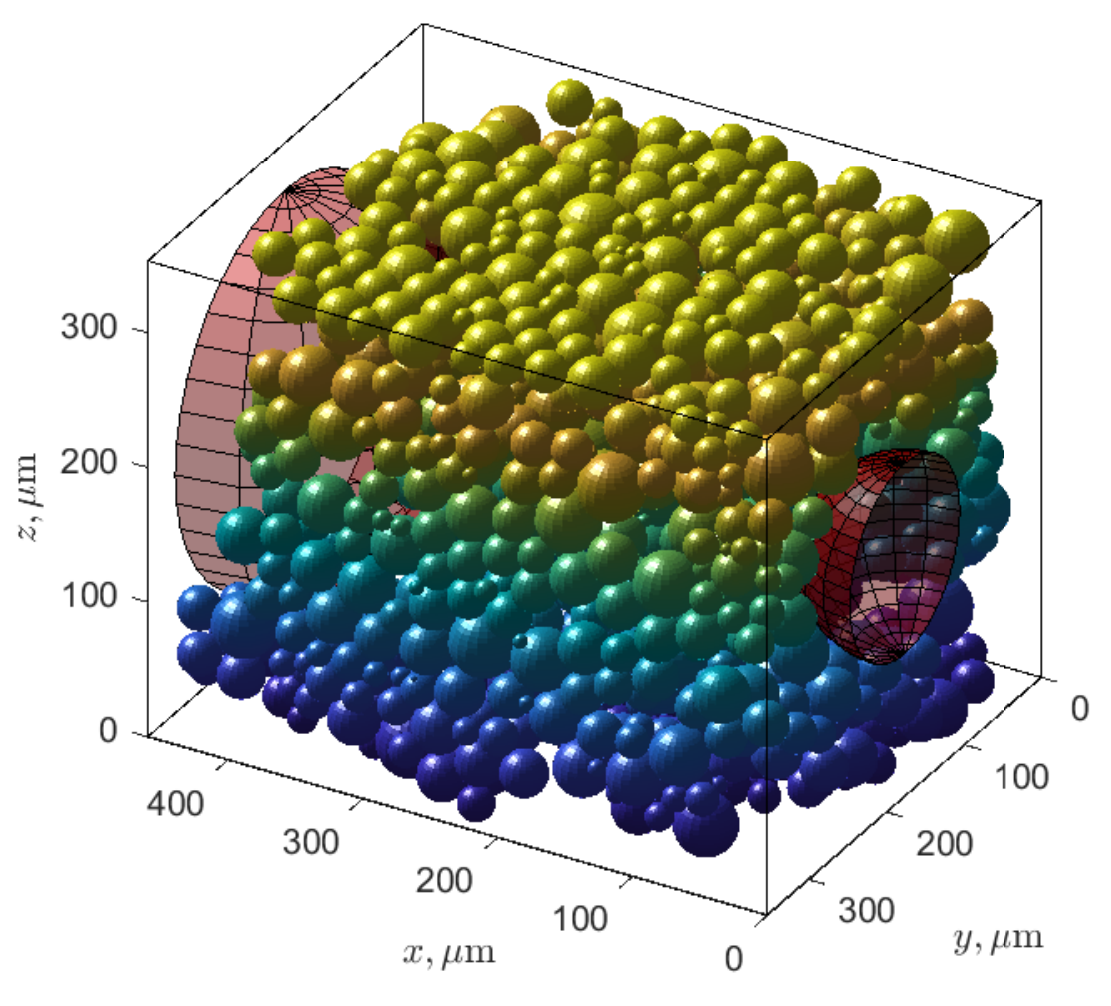}}
\end{subfigure}
\\
\begin{subfigure}[b]{.5\textwidth}
  \centering
  \subcaptionbox{\label{fig:Eg3:all:C}}{\includegraphics[width=0.95\textwidth]{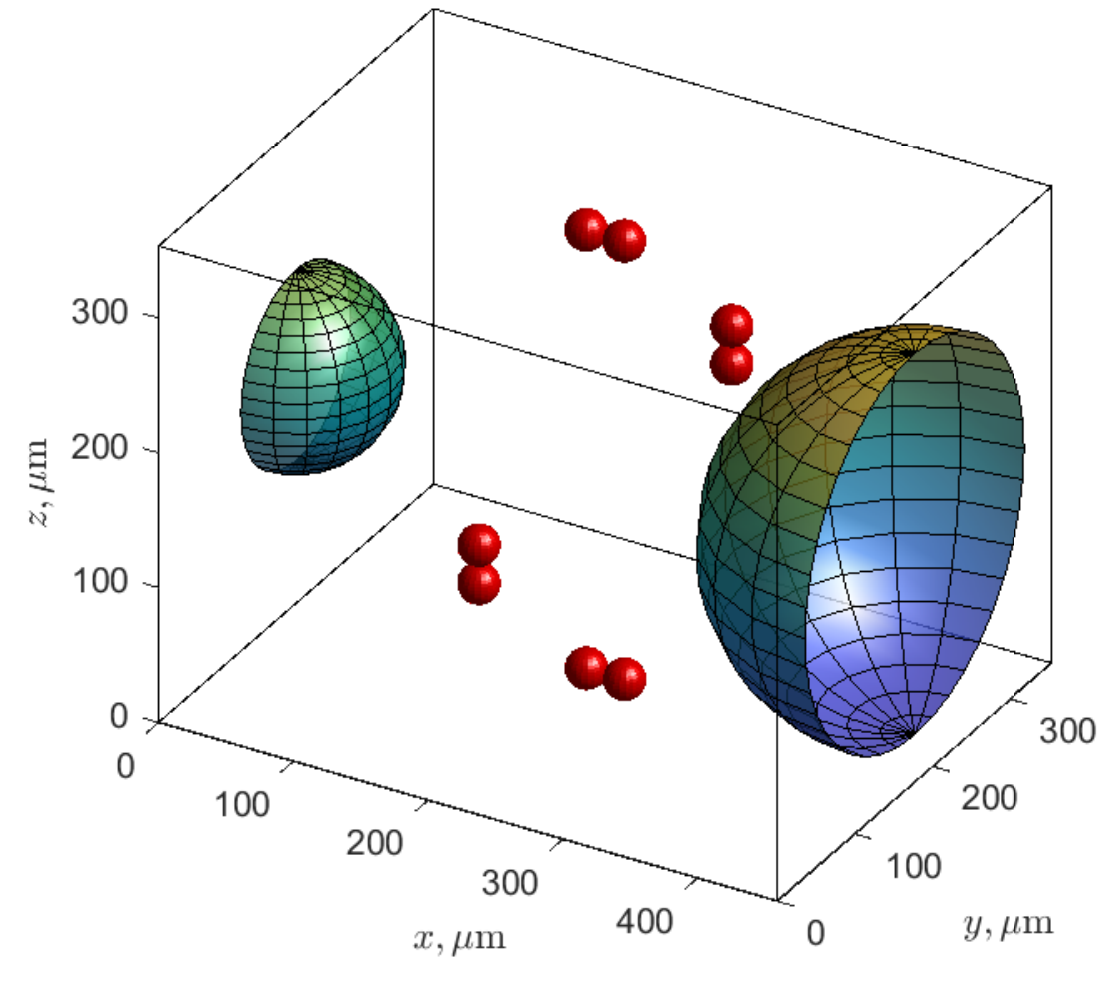}}
\end{subfigure}
\begin{subfigure}[b]{.5\textwidth}
  \centering
  \subcaptionbox{\label{fig:Eg3:all:D}}{\includegraphics[width=0.95\textwidth]{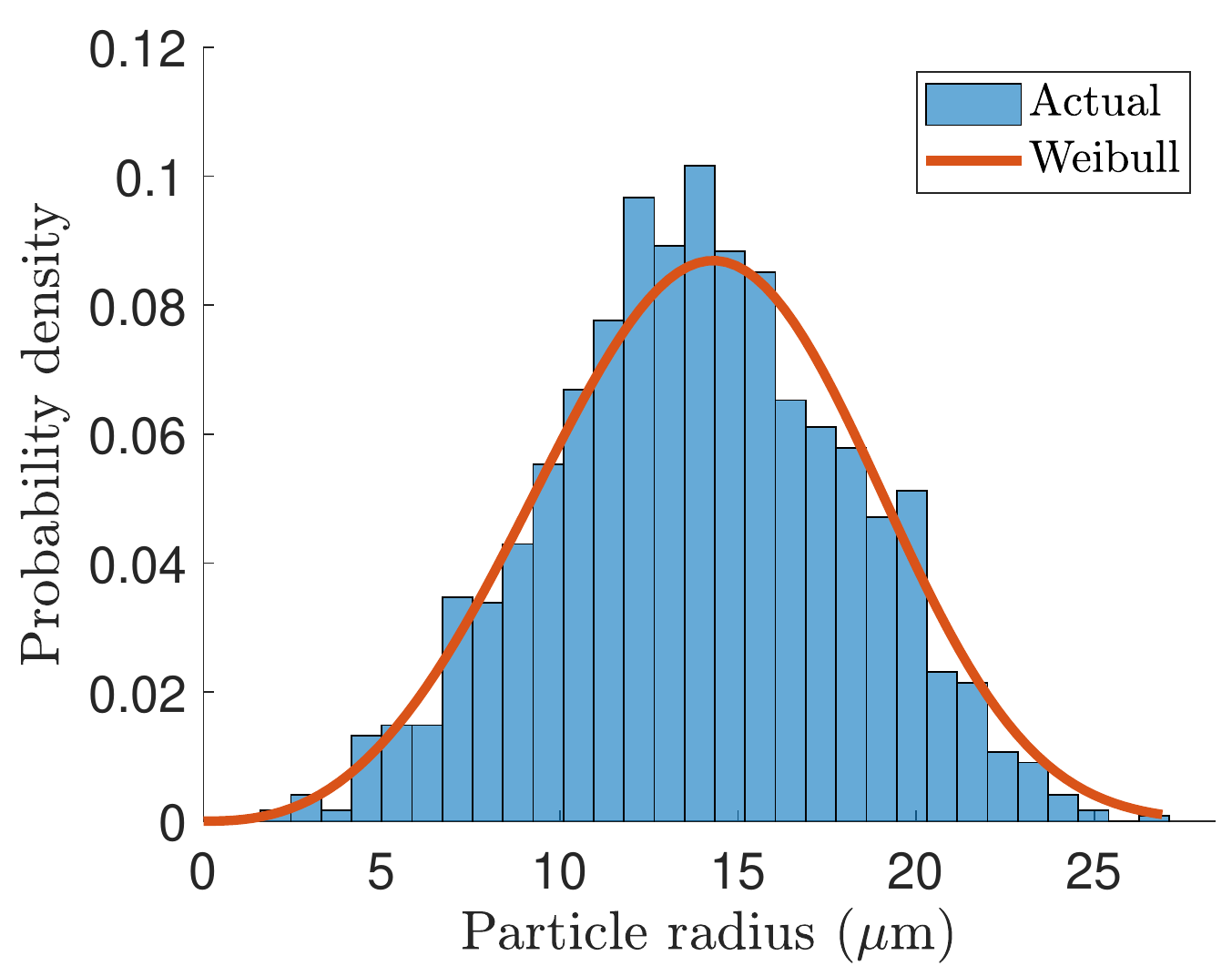}}
\end{subfigure}
\caption{\footnotesize
(a) Example 3A: a unit cube with side length \eqref{eq:std:len:eg3} minus two equal hemispheres, filled with random spheres and Weibull distribution \eqref{Weib}, \eqref{Weib:params}. (b) Example 3B: a similar computation with a non-cubic domain minus two non-equal hemispheres.
(c) The computational domain and first parent sphere parents for Example 3B. (d) Actual sphere radius distribution in Example 3B, compared to the given Weibull distribution.
}
\label{fig:Eg3:all}
\end{figure}

\medskip \noindent{\textbf{Example 3A.}} In the first sample run, we use the Weibull distribution \eqref{Weib}, \eqref{Weib:params} to fill a domain $\mathcal{V}$
based on a cube-shape with side length
\beq\label{eq:std:len:eg3}
\verb|std_length|=30\,\bar{r}
\eeq
(for Weibull distribution, $\bar{r}$ is given by \eqref{Weib:mean}), with the subtraction of two hemispheres of equal radii $0.5\times\verb|std_length|$. A script \verb|Example3A_Cube_Hemisph_Generate_and_Plot.m| (Appendix \ref{appendix:eg3:generate}) calls the sphere placing routine \verb|Example3GenerateSpheres.m| with the following parameters.
\begin{itemize} \setlength\itemsep{0ex}
  \item \verb|ProbabilityDistr|: Weibull \eqref{Weib}, \eqref{Weib:params}.
  \item \verb|BrickSideLengths|: \verb|[1; 1; 1]*std_length|.
  \item \verb|HemisphereRadii|: \verb|[0.5; 0.5]*std_length|.
  \item \verb|FaceGoal|: 0.4.
  \item \verb|BodyGoal|: 0.4.
  \item \verb|SphereContactParameter|: 0.1.
\end{itemize}
This example took 49 minutes to complete in the workstation configuration (138 minutes in the laptop configuration). The resulting spherical arrangement is shown in Figure~\ref{fig:Eg3:all}\,(a).
%Lomonosov: 49; laptop: 138

\medskip \noindent{\textbf{Example 3B.}} In the second sample run, the same Weibull distribution \eqref{Weib}, \eqref{Weib:params} and the typical domain size \eqref{eq:std:len:eg3} are used to fill a domain based on a non-cubical brick,
with subtraction of two hemispheres of non-equal radii. A script based on \verb|Example3A_Brick_Hemisph_Generate_and_Plot.m| of Example 3A calls the sphere placing routine with different domain and computation parameters, as follows.
\begin{itemize} \setlength\itemsep{0ex}
  \item \verb|ProbabilityDistr|: Weibull \eqref{Weib}, \eqref{Weib:params}.
  \item \verb|BrickSideLengths|: \verb|[1.3; 1; 1]*std_length|.
  \item \verb|HemisphereRadii|: \verb|[0.2; 0.4]*std_length|.
  \item \verb|FaceGoal|: 0.4.
  \item \verb|BodyGoal|: 0.4.
  \item \verb|SphereContactParameter|: 0.2.
\end{itemize}
The resulting spherical filling and related graphs is shown in Figure~\ref{fig:Eg3:all}\,(b,c,d). This computation took 41 minutes to complete in the workstation configuration (144 minutes in the laptop configuration). The resulting sphere size distribution histogram (Fig.~\ref{fig:Eg3:all}\,(d)) shows a good agreement with the given probability density function.
% 41 on Lomonosov; 144 on laptop

%% ------TB:-------
%This is an example of filling a space which has spherical boundaries. This is accomplished by placing a sphere on the $x_{\mathrm{min}}$ face and the $x_{\mathrm{max}}$ face. An initial set of parents is required to build off of, so a pair of spheres is placed on the $y_{\mathrm{min}}, y_{\mathrm{max}}, z_{\mathrm{min}}, z_{\mathrm{max}}$ faces. \textit{scale\_input} = 31.4, \textit{shape\_input} = 3.55 are parameters for the Weibull distribution. \textit{x\_max\_input}, \textit{y\_max\_input}, \textit{z\_max\_input} = 20 determine the dimensions of the rectangle to be filled. \textit{sphere\_rad\_1\_input} = 10 determines the radius of the spherical boundary on the $x_{\mathrm{min}}$ face. \textit{sphere\_rad\_2\_input} = 10 determines the radius of the spherical boundary on the $x_{\mathrm{max}}$ face.
%
%This example took 748 seconds to complete.

%-------------------------------------

\section{A physical example: bonding modeling in powder bed 3D printing process} \label{egPhys}

% Powder bed generated and plotted by Method 2 in new notation: Example4_use_Method2_to_generate_powder_bed.m
% Got:
%  - Data: Example4_powder_bed.mat
%  - Figure of powder bed: Example4_powder_bed.fig, .pdf

% Leave travis' pictures because they are all correct! Right sizes, right radii... all good! In our folder Example4laser,

We now use the sphere packing Method 2 to model heat-based bonding in the additive manufacturing process that uses a metal powder bed and a guided laser beam to heat the spherical particles and thus form bonds between them, as described in Ref.~\cite{spierings2009comparison}. Particle size distribution for 316L stainless steel powder is approximated by the Weibull distribution \eqref{Weib}, \eqref{Weib:params} \cite{SteubenJohnC2016Demo,spierings2011influence}.

\subsection{The heat source and powder bed models}\label{egPhys:physics}

For the heat source model, physical assumptions are as follows.

\begin{itemize}\setlength\itemsep{0.5ex}

\item The heat flux from laser into the $i^{\rm th}$ spherical particle is given by
\[
q_{i_{\rm laser}} = Q\dfrac{{r_i}^3}{{r_\ell}^3}\, ,
\]
where $Q$ is the total power of the laser, $r_i$ is the radius of the particle, and $r_\ell$ is the radius of the laser beam.

\item
The heat flux from convection is
\[
q_{i_{\rm conv}} = k_b(T_R - T_i)\, ,
\]
where $k_b$ is the heat transfer coefficient, and $T_R$ is the temperature of the surrounding air.

\item The heat flux between two particles is given by
\[
q_{ij} = k_t\Big( T_j  - T_i\Big),
\]
where $k_t$ is the thermal conductivity, and $T_i$, $T_j$ denote the temperatures of particles $i$ and $j$.
The total heat flux into the $i^{\rm th}$ particle thus becomes
\[
q_i = q_{i_{\rm laser}} + q_{i_{\rm conv}} + \sum\limits_{j = 1}^{N} q_{ij}.
\]

\item Using the discrete time stepping $t_1$, $t_2=t_1+\Delta t$, \ldots, the temperature update for the $i^{\rm th}$ particle is expressed by
\[
T_{i}^{t + \Delta t} = T_i^t + \frac{q_i^t}{m_i C_p}\Delta t,
\]
where $T_i^t$ is the particle's temperature at the previous time step, $T_{i}^{t + \Delta t}$ is the temperature at the following time step, $q_i^t$ is the total initial energy flux, $m_i$ is the mass of the $i^{\rm th}$ spherical particle, and $C_p$ denotes the material specific heat.

\item If particles $i$ and $j$ are in contact, and both above the sintering temperature $T_s$,  a bond is formed between them.

\end{itemize}

The following sample values for the simulated print using the parameters below were adapted from Ref.~\cite{SteubenJohnC2016Demo}.

\begin{itemize} \setlength\itemsep{0.5ex}
    \item Total power of the laser: $Q = 100~\mathrm{W}$.

    \item Thermal conductivity of air: $\tau_a = 0.262~\mathrm{W}/(\mathrm{m \cdot K})$.

    \item Thermal conductivity of steel: $\tau_s = 15~{\mathrm{W}}/(\mathrm{m\cdot  K})$.

    \item Heat transfer coefficient at the air boundary: $k_b = {\pi \tau_a r_{\mathrm{avg}}}/{2} = 2.9090 \times 10^{-6}~ {\mathrm{W}}/{\mathrm{ K}}$.

    \item Heat transfer coefficient of steel: $k_t = {\pi \tau_s r_{\mathrm{avg}}}/2 = 3.3309 \times 10^{-4}~{\mathrm{W}}/{\mathrm{K}}$.

    \item Specific heat capacity of steel $C_p = 0.5~{\mathrm{J}}/({\mathrm{g \cdot K}})$.

    \item Steel density $\rho = 8 \times 10^{-12}~{\mathrm g}/({\mu\mathrm m})^3$.

    \item Mass of a spherical particle $m_i = ({4\pi}/{3})\,\rho\, r_i^3$.

    \item Laser radius $r_{\ell} = 50\,\mathrm{\mu m}$.

    \item Ambient air temperature: $T_R = 300~\mathrm{K}$. %*** needs to be 100C ? check?

    \item Sintering temperature: $T_s = 1000 ~\mathrm{K}$. %in fact, from the paper, should be 1000 C !
\end{itemize}

\subsection{A simulated print of a square}\label{egPhys:sim}

\begin{figure}[htbp]
\begin{subfigure}[b]{.5\textwidth}
  \centering
  \subcaptionbox{\label{fig:Eg:Laser:init}}{\includegraphics[width=0.85\textwidth]{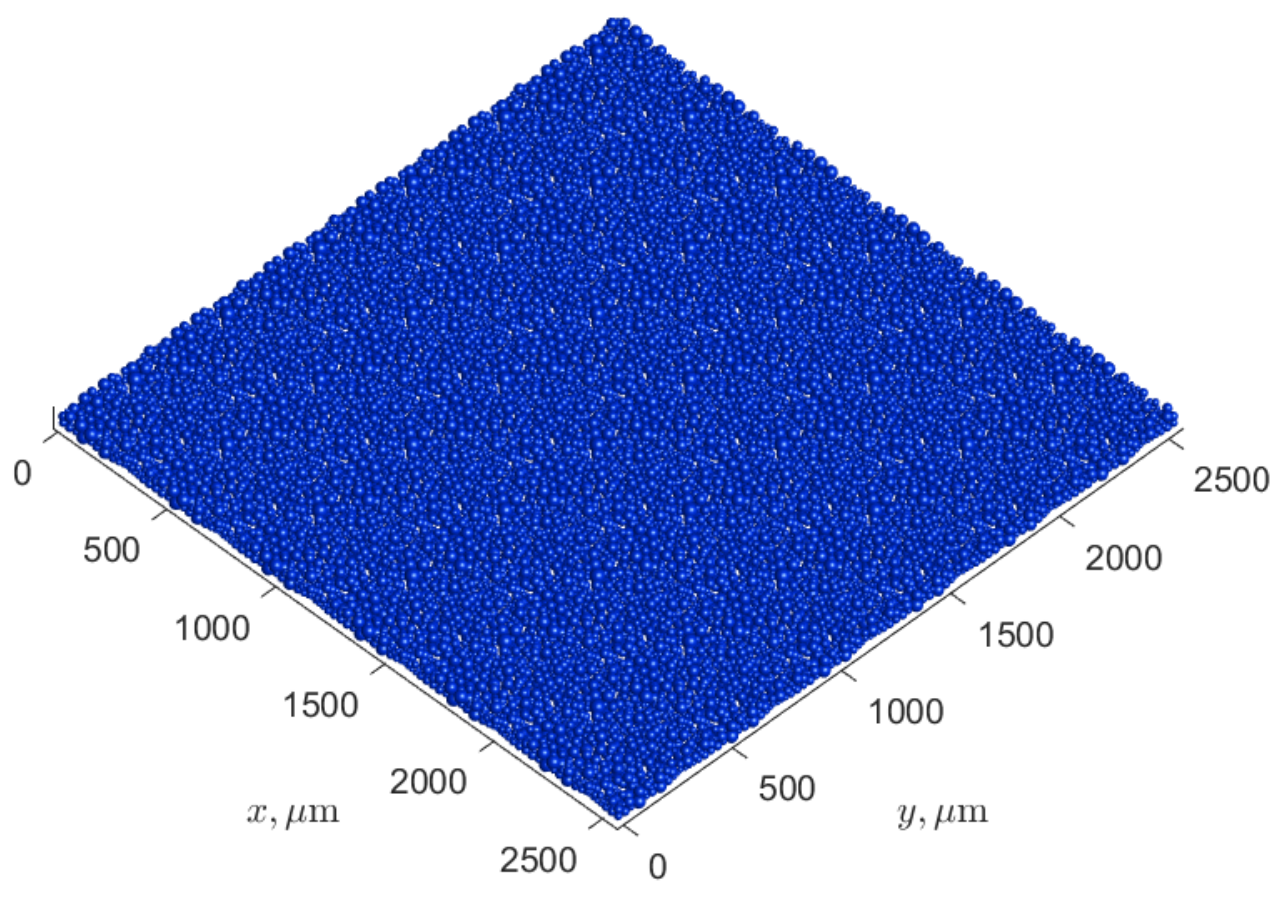}}
\end{subfigure}
\begin{subfigure}[b]{.5\textwidth}
  \centering
  \subcaptionbox{\label{fig:Eg:Laser:path}}{\includegraphics[width=0.85\textwidth]{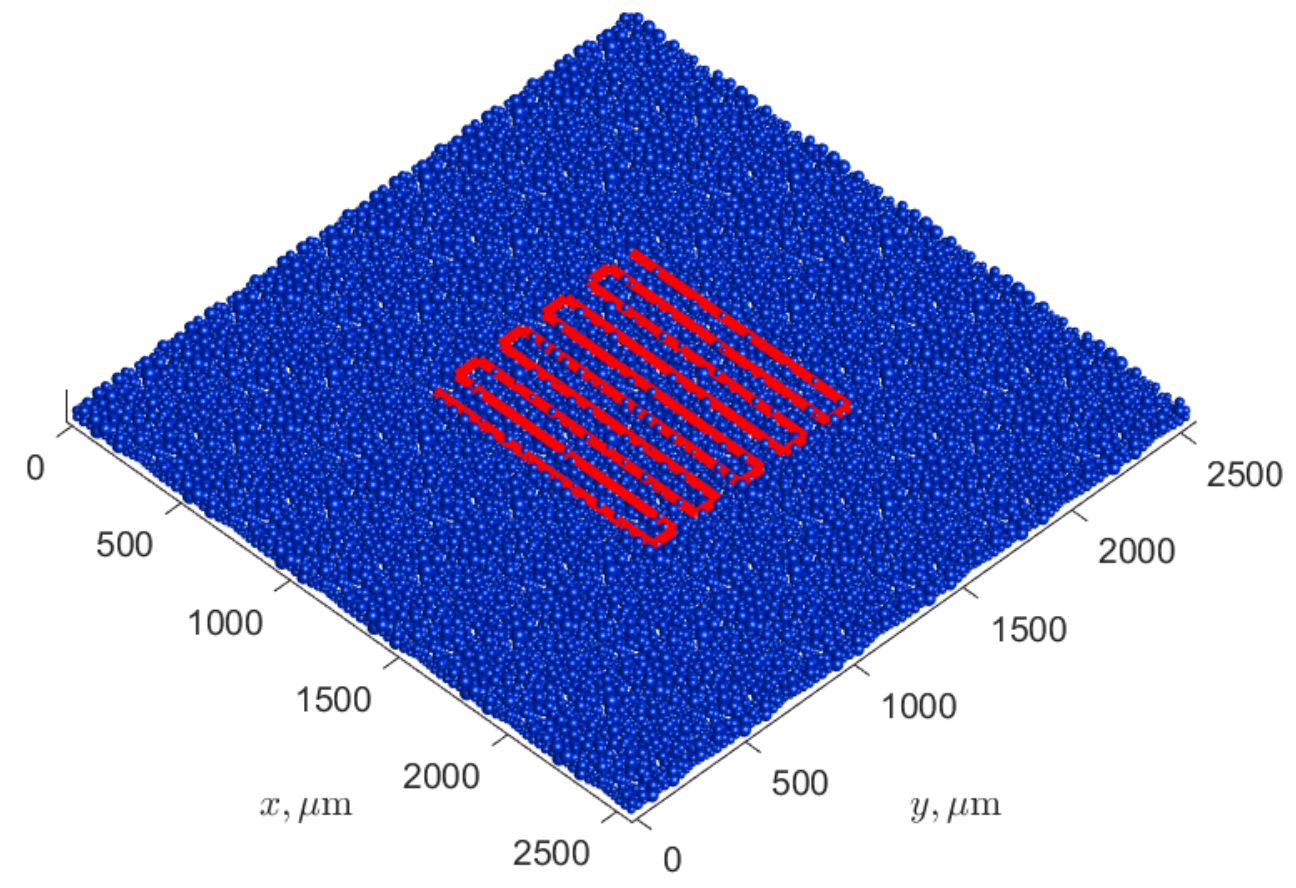}}
\end{subfigure}
\\
\begin{subfigure}[b]{.5\textwidth}
  \centering
  \subcaptionbox{\label{fig:Eg:Laser:pass1}}{\includegraphics[width=0.95\textwidth]{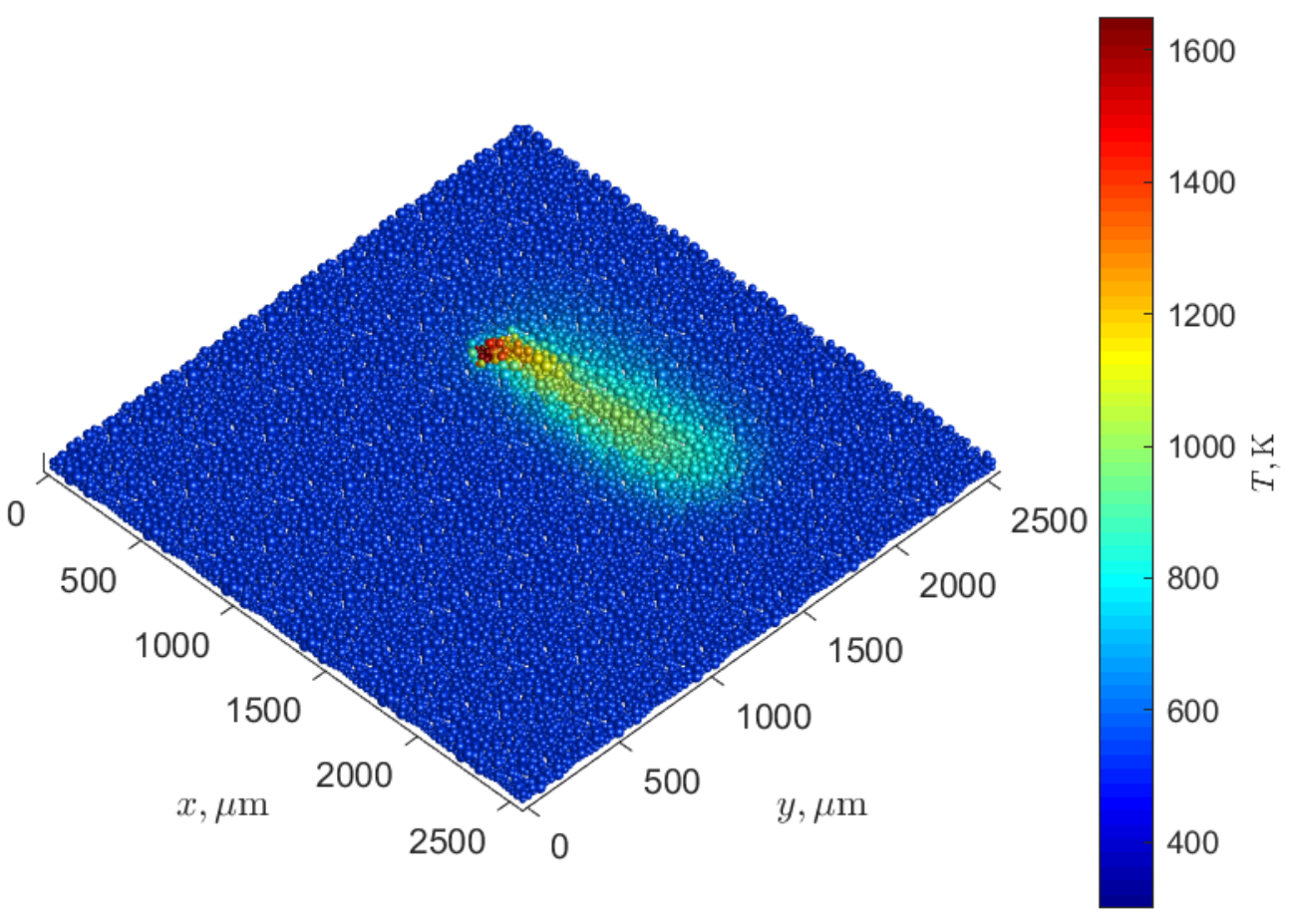}}
\end{subfigure}
\begin{subfigure}[b]{.5\textwidth}
  \centering
  \subcaptionbox{\label{fig:Eg:Laser:pass2}}{\includegraphics[width=0.95\textwidth]{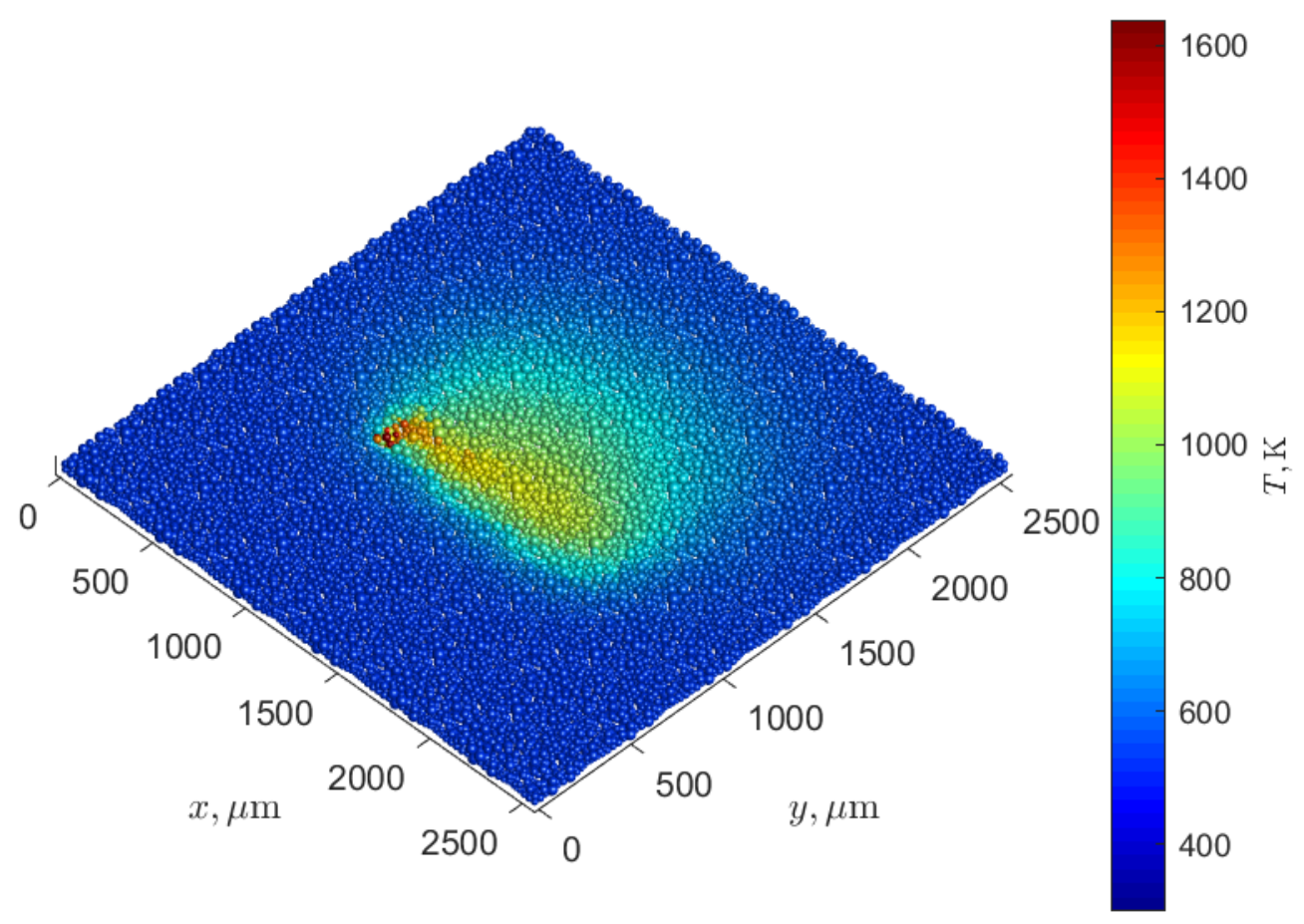}}
\end{subfigure}
\caption{\footnotesize
A simulated print of a square (Section \ref{egPhys}). (a) The printing domain (size in $\text{\textmu m}$). (b) The laser beam path. (c) Second pass of the laser (temperature scale in Kelvins). (d) Final pass of the laser.
}
\label{fig:Eg:Laser}
\end{figure}

\begin{figure}[htbp]
\begin{subfigure}[c]{.55\textwidth}
  \centering
  \subcaptionbox{\label{fig:Eg:Laser:init}}{\includegraphics[width=0.95\textwidth]{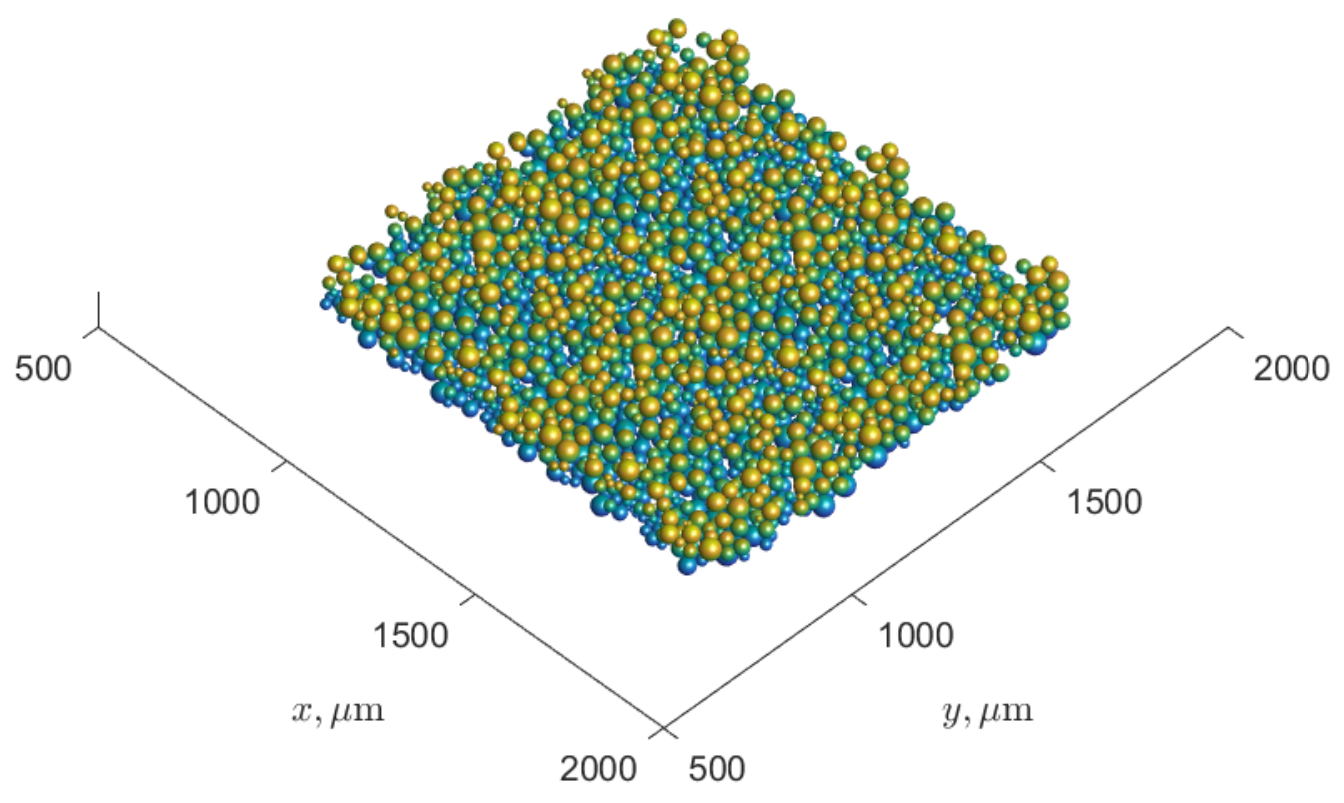}}
\end{subfigure}
\begin{subfigure}[c]{.45\textwidth}
  \centering
  \subcaptionbox{\label{fig:Eg:Laser:path}}{\includegraphics[width=0.95\textwidth]{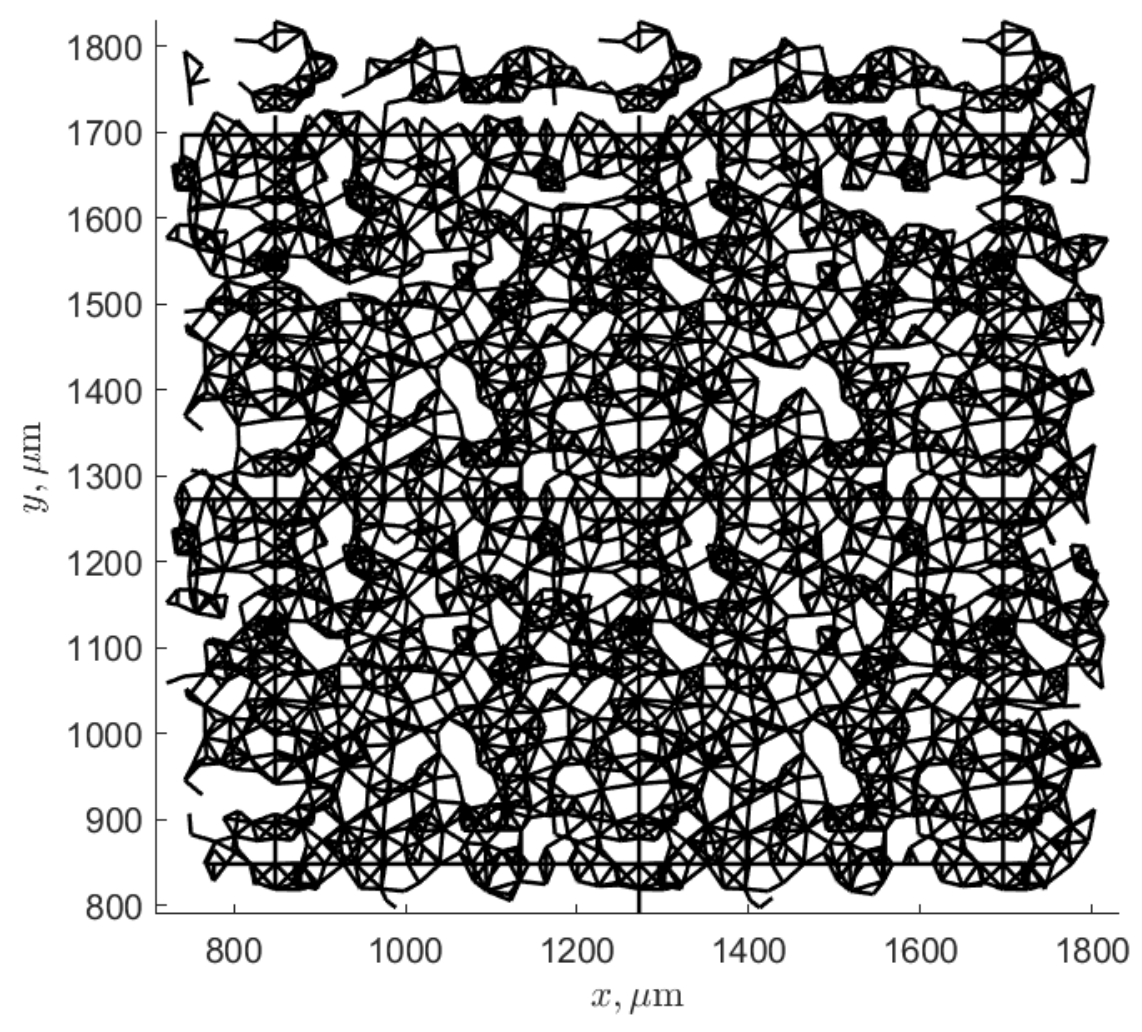}}
\end{subfigure}
\caption{\footnotesize
The simulated print of a square (Section \ref{egPhys}). (a) Particles bonded during the simulated print (b) Bonds formed between particles.
}
\label{fig:Eg:Laser2}
\end{figure}

A sample powder bed was generated using Method 2 (Section \ref{sec:FullMethod2}) with parameters
\begin{itemize}\setlength\itemsep{0ex}
  \item \verb|ProbabilityDistr|: Weibull \eqref{Weib}, \eqref{Weib:params},
  \item \verb|FaceGoal|: 1.0,
  \item \verb|BodyGoal|: 0.9,
  \item \verb|SphereContactParameter|: 0.1,
  \item \verb|ParentParameter|: 0.5,
  \item \verb|BrickSideLengths|: \verb|[1; 1; 0.12] * std_length|, where $\verb|std_length|=30\,\bar{r}$,
  \item \verb|BrickNumbers|: \verb|[6; 6; 1]|.
\end{itemize}
The computation took around five minutes, yielding the domain of size $\sim 2545\times 2545\times 50.89~\text{\textmu m}$, with the average particle radius $14.14~\text{\textmu m}$ (Figure \ref{fig:Eg:Laser}\,(a)). A laser path shown in Figure \ref{fig:Eg:Laser}\,(b) was chosen to simulate the print of a small square in the middle of the domain by building thermally induced connections between the particles. Figures \ref{fig:Eg:Laser}\,(c,d) show the temperature maps on the second and the final laser pass. Figure \ref{fig:Eg:Laser2} depicts particles that were bonded in the result of the simulated print process and the top view of the bond graph.

%==================Summary ======================
\section{Discussion} \label{sec:Discuss}

The current manuscript describes a \verb|Matlab| implementation of a geometry-based filling of a three-dimensional domain with spheres whose radii are randomly distributed according to a given probability distribution (an arbitrary probability distribution supported by \verb|Matlab|). Two methods, Method 1 (Section \ref{sec:FullMethod1}) and Method 2 (Section \ref{sec:FullMethod1}), fill a parallelepiped-shaped domain with spheres, using one or more copies of a standard ``unit brick" formed with user-prescribed sizes. The two methods are implemented as two freely interchangeable \verb|Matlab| functions \verb|Method1GenerateSpheres.m| and \verb|Method2GenerateSpheres.m| with identical input and output parameters. The first method produces a less symmetric unit brick having different random filling of edges and faces, while the second method copies three initially formed edges and faces to opposite edges and faces. The first method places the spheres so that they touch all faces of the unit brick, and uses reflections to make its copies to fill out the given domain. The second method places centers of the random spheres on the faces, and multiplies unit bricks to tile the total volume by simple copying. Both methods record and return particle radii, positions, and pairwise contacts.

Both presented methods are based on simple geometry of inter-particle contacts, in particular, geometrical determination of optimal ``parent" pairs for each upcoming random-sized particle (as opposed to, for example, more complex physical sphere filling methods based on mechanical interactions and resulting particle dynamics and settling).

Methods 1 and 2 provide similar (but different) sphere fillings, with similar run times. In both methods, computations run with the aim of reaching the user-set \verb|FaceGoal| and \verb|BodyGoal| parameters, representing the area and volume fraction of the spheres to the unit brick domain. Both methods work much faster with smaller unit bricks; however, this affects the actual probability distribution of the spheres. Indeed, as the unit brick gets smaller, in order to achieve face and body filling goals, spheres that do not fit may be discarded, and hence the probability distribution of the spherical radii may be more skewed towards smaller radii. However, even for rather small unit bricks (size $\sim 30$ average particle radii), the probability histograms appear to agree well with the theoretical probability density function.

Three run examples of Method 1 are presented in Section \ref{sec:FullMethod1:eg}. Examples 1A and 1B fill a unit cube with different side lengths $15\,\bar{r}$ and $30\,\bar{r}$ in terms of the average sphere radius $\bar{r}$ \eqref{Weib:mean} of the Weibull distribution \eqref{Weib}. A sample multi-cube domain is built, and the connectivity graph joining contacting particles is presented (Figure \ref{fig:Method1:2:Full:conn}). In Example 1C, a non-cubic unit brick is generated, and the total domain is filled by joining $4\times 4\times 2$ unit bricks.
An example of the use of Method 2 is presented in Section \ref{sec:FullMethod2:eg} where the Gamma distribution is used to pick random spheres. Run times are compared depending on the relative unit brick sizes and the values of the sphere contact parameter.

Example 3 (Section \ref{sec:eg3:hemisph}) features an adaptation of the sphere filling algorithm to a more complex-shaped domain: a parallelepiped with two hemispheres removed. The latter can have arbitrary (reasonably small) radii, and are centered in the middles of two opposite faces $x=0$ and the maximal $x$. Two run examples demonstrate successful fillings of cubic and non cubic domains, with equal and non-equal removed hemispheres.

Section \ref{egPhys} presents an application of spherical filling to model the creation of heat-induced connections between discrete metal particles in powder bed 3D printing process, where a laser beam is employed to locally heat the powder bed above the sintering temperature. Temperature snapshots, the set of bonded particles, and the corresponding connectivity graph are presented for a sample simulation involving five laser passes.

In both Methods 1 and 2 presented in this work, computations that generate one unit brick are essentially based on for-loops that cannot be parallelized in a straightforward manner. The main idea behind the creation of these routines was the simplicity of coding and use, while parallelization would require a significant conceptual rethinking. In the case when the total domain is obtained by joining together more than one unit brick, the particle contacts and the size distribution are preserved, while the domain acquires an effective artificial ``frame" made of shared or reflected  cube edges and faces. This illustrates the tradeoff between the computational time and obtaining a better looking sphere-filled domain with less or no symmetry.

The main goal of the present work was the development of a versatile code that can be quickly and efficiently used to generate spherical fillings of domains of different size and shape, with an ability to choose a probability distribution from those available in \verb|Matlab| and tune the distribution parameters, without the need to change the routines themselves. This qualities make the current algorithms rather different from those that generate random or non-random close packings of spheres (or other objects) of the same size (e.g., \cite{torquato2000random, williams2003random}). A common measure of optimality of such designs is the volume fraction $\phi$ occupied by the repeated object. It has been shown that random close packings of identical spheres can attain $\phi\sim 0.64$, whereas the face-centered cubic lattice corresponds to $\phi\sim 0.74$ (\cite{torquato2000random} and references therein). The code presented in the current work, however, aims essentially at problems where spherical radii are unequal and follow some given nonsingular (usually continuous) probability distribution that models a physical situation. The volume fraction occupied by spheres would significantly depend on the chosen probability distribution and its specific parameters, as well as on the domain shape (unless its dimensions are large compared to $\bar{r}$). Since both computational methods presented here require an \emph{a priori} specification of the desired filled volume fraction, it would require some kind of shooting method to determine a close-to-maximal volume fractions for which computations can be finished in reasonable time, separately for each setup of interest. The examples considered above were based on practically determined volume fractions, realistically attainable for the general code and the respective distributions. Further details regarding the properties of packings generated by the presented code would require additional studies that are out of scope of this work, but indeed constitute an interesting possible future work direction.

The authors realize that the current (open-source) implementation can be further improved and generalized, which an interested reader is warmly invited to do. In particular, the presented methods in their current form do not provide solutions to optimal packing problems, which is definitely an interesting avenue of experimentation and generalization of these algorithms.

\subsubsection*{Acknowledgements}

The authors are grateful to the anonymous referees for valuable suggestions and references, and to NSERC of Canada for support through the Discovery grant RGPIN-2019-05570 and a USRA fellowship.

{\footnotesize
\bibliography{bibliography19c}

\begin{thebibliography}{10}

\bibitem{WangJie1999PoUS}
J.~Wang, ``Packing of unequal spheres and automated radiosurgical treatment
  planning,'' {\em Journal of Combinatorial Optimization}, vol.~3, no.~4,
  pp.~453--463, 1999.

\bibitem{michopoulos2018multiphysics}
J.~G. Michopoulos, A.~P. Iliopoulos, J.~C. Steuben, A.~J. Birnbaum, and S.~G.
  Lambrakos, ``On the multiphysics modeling challenges for metal additive
  manufacturing processes,'' {\em Additive Manufacturing}, vol.~22,
  pp.~784--799, 2018.

\bibitem{XIN2018373}
H.~Xin, W.~Sun, and J.~Fish, ``Discrete element simulations of powder-bed
  sintering-based additive manufacturing,'' {\em International Journal of
  Mechanical Sciences}, vol.~149, pp.~373--392, 2018.

\bibitem{HIFI2019482}
M.~Hifi and L.~Yousef, ``A local search-based method for sphere packing
  problems,'' {\em European Journal of Operational Research}, vol.~274, no.~2,
  pp.~482--500, 2019.

\bibitem{StoyanYu.2016PUSi}
Y.~Stoyan, G.~Scheithauer, and G.~Yaskov, ``Packing unequal spheres into
  various containers,'' {\em Cybernetics and Systems Analysis}, vol.~52, no.~3,
  pp.~419--426, 2016.

\bibitem{torquato2000random}
S.~Torquato, T.~M. Truskett, and P.~G. Debenedetti, ``Is random close packing
  of spheres well defined?,'' {\em Physical review letters}, vol.~84, no.~10,
  p.~2064, 2000.

\bibitem{williams2003random}
S.~Williams and A.~Philipse, ``Random packings of spheres and spherocylinders
  simulated by mechanical contraction,'' {\em Physical Review E}, vol.~67,
  no.~5, p.~051301, 2003.

\bibitem{spierings2009comparison}
A.~B. Spierings and G.~Levy, ``Comparison of density of stainless steel {316L}
  parts produced with selective laser melting using different powder grades,''
  in {\em Proceedings of the Annual International Solid Freeform Fabrication
  Symposium}, pp.~342--353, Austin, TX, 2009.

\bibitem{SteubenJohnC2016Demo}
J.~C. Steuben, A.~P. Iliopoulos, and J.~G. Michopoulos, ``Discrete element
  modeling of particle-based additive manufacturing processes,'' {\em Computer
  Methods in Applied Mechanics and Engineering}, vol.~305, pp.~537--561, 2016.

\bibitem{spierings2011influence}
A.~B. Spierings, N.~Herres, and G.~Levy, ``Influence of the particle size
  distribution on surface quality and mechanical properties in {AM} steel
  parts,'' {\em Rapid Prototyping Journal}, 2011.

\end{thebibliography}
\bibliographystyle{ieeetr}
}

%===================Appendices===========================
\begin{appendix}

\section{Data generation and plotting script for sphere packing Method 1, Example 1A } \label{appendix:eg1:generate}

\begin{itemize}\setlength\itemsep{0.5ex}
  \item Main file: \verb|Example1A_Method1_Generate_and_Plot.m|
  \item Output:
  \begin{itemize}\setlength\itemsep{0ex}
    \item Figure \ref{fig:Method1:all} (b,d,f).
    \item Figure similar to Figure \ref{fig:Method1:2:Full:conn} (a,b) for Example 1B.
    \item \verb|Matlab| data file \verb|Example1A_Method1_Results.mat|
  \end{itemize}
  \item Required additional files (in the same folder):
  \begin{itemize}\setlength\itemsep{0ex}
    \item \verb|Method1GenerateSpheres.m|: the main Method 1 sphere filling routine
    \item Auxiliary routine files for \verb|Method1GenerateSpheres.m|:
    \begin{itemize}\setlength\itemsep{0ex}
      \item \verb|M1Position2Xmax.m|
      \item \verb|M1Position2Xmin.m|
      \item \verb|M1Position2Ymax.m|
      \item \verb|M1Position2Ymin.m|
      \item \verb|M1Position2Zmax.m|
      \item \verb|M1Position2Zmin.m|
      \item \verb|M1Position3.m|
      \item \verb|M1Search2D.m|
      \item \verb|M1Search3D.m|
    \end{itemize}
  \end{itemize}
\end{itemize}

The description of some commands and run parameters used in the script is as follows.
\begin{itemize}\setlength\itemsep{0ex}
  \item \verb|rng(0)|:  Set \verb|Matlab| pseudorandom seed to zero (default).
  \item \verb|Weibull_scale|, \verb|Weibull_shape|: the distribution parameters $\lambda$, $k$ in \eqref{Weib:params}.
  \item \verb|ProbabilityDistribution|: the \verb|Matlab| distribution variable for the PDF \eqref{Weib}, \eqref{Weib:params}.
  \item \verb|cube_side_length|: side length of a unit cube; here 15 times the average sphere radius $\bar{r}$.
  \item \verb|FaceGoal|, \verb|BodyGoal|: minimal percentages of area and volume fill ratios of cube sides. Here 0.8 and 0.55.
  \item \verb|BrickSideLengths = [1;1;1]* std_length| where $\verb|std_length|=15\,\bar{r}$.
  \item \verb|BrickNumbers = [2;2;1]|: numbers of unit bricks in $x$, $y$, $z$ directions, making up the total domain $\mathcal{V}$.
  \item \verb|SphereContactParameter|, \verb|ParentParameter|: the contact and the parent parameter (see Section \ref{sec:FullMethod1:funct}). Here 0.2 and 0.5.
\end{itemize}

\begin{lstlisting}[language=Matlab]
%% Matlab script for Example 1A: Method 1, using Weibull distribution
% Produces the sphere filling of the domain V and plots of Figure 2 (a, c, e) and figures similar to Fig.3 (a,b) 
% Required files:
% * Method1GenerateSpheres.m -- main sphere filling routine
% * Auxiliary routine files for Method1GenerateSpheres.m :
%    - M1Position2Xmax.m
%    - M1Position2Xmin.m
%    - M1Position2Ymax.m
%    - M1Position2Ymin.m
%    - M1Position2Zmax.m
%    - M1Position2Zmin.m
%    - M1Position3.m
%    - M1Search2D.m
%    - M1Search3D.m

clear all; close all; clc;

%% Sphere Packing parameters: Weibull distribution

% -- use Weibull distribution implemented in Matlab
% with parameters from Steuben, Iliopoulos, Michopoulos (2016)

rng(0); %random seed zero: Matlab default
Weibull_scale = 31.4/2; % scale parameter lambda for the radius in Weibull distribution; in micrometers. The scale parameter for *diameter* is 31.4.
Weibull_shape = 3.55; % shape parameter k in Weibull distribution; dimensionless

ProbabilityDistribution = makedist('Weibull','a',Weibull_scale ,'b', Weibull_shape);

average_radius = mean(ProbabilityDistribution);
FaceGoal = 0.8;  % desired sphere/face area fraction; recommended for Method 1: 0.8
BodyGoal = 0.55; % desired sphere/volume fraction; recommended for Method 1: 0.55

%Unit brick: here unit cube
std_length = 15*average_radius; 
BrickSideLengths = [1;1;1]* std_length;

%numbers of cubes in x, y, z directions to fill the total domain V
BrickNumbers = [2;2;1];

SphereContactParameter = 0.2; %This is the contact parameter, within [0,1]. If particles are within SphereContactParameter*average_radius of each other, they are considered to be in contact
ParentParameter = 0.5; %This is the parent parameter, within [0,1]. If particles are within ParentParameter*average_radius of each other, they are considered to be potential parents

%% Run the sphere generation Method 1 algorithm and save results.

tic;
[FinalNSpheres, UnitBrickNSpheres, Positions, Radii, Contacts, ListXmin, ListYmin, ListZmin, ListXmax, ListYmax, ListZmax] = ...
  Method1GenerateSpheres( ...
    ProbabilityDistribution, ...
    FaceGoal, BodyGoal, ...
    SphereContactParameter, ParentParameter, ...
    BrickSideLengths, BrickNumbers );

sphere_filling_total_time = toc

%Save computations 
save('Example1A_Method1_Results.mat')

%% Now some plots: reproduce Fig. 

[x_sph, y_sph, z_sph] = sphere;

%% Unit Cube Plot
figure(11)
hold on
axis equal
percent = 0;
List = 1:UnitBrickNSpheres; %particles you want to display
light               % create a light
lighting gouraud    % preferred method for lighting curved surfaces
for count = 1:size(List,2)
    i = List(count);
    surf(Radii(i)*x_sph + Positions(1,i), Radii(i)*y_sph + Positions(2,i), Radii(i)*z_sph + Positions(3,i),'EdgeColor','none','FaceLighting','gouraud')
    
    if mod(count,1000)==0
        percent = 100*(count/UnitBrickNSpheres); %Keeping track of progress
        disp(['Plotting Unit Cube ',num2str(percent), '% complete'])
    end
    
end
xlim([0, BrickSideLengths(1)]);
ylim([0, BrickSideLengths(1)]);
zlim([0, BrickSideLengths(1)]);
xlabel('$x, {\rm \mu m}$', 'Interpreter', 'latex');
ylabel('$y, {\rm \mu m}$', 'Interpreter', 'latex'); 
zlabel('$z, {\rm \mu m}$', 'Interpreter', 'latex');
view(60, 45);
hold off;

% Save figure - uncomment as needed
% savefig('Example1A_Method1_UnitCube.fig');
% print('Example1A_Method1_UnitCube','-dpdf');

%% Unit Cube Middle Third
figure(13)
hold on
axis equal
List = find((Positions(3,1:UnitBrickNSpheres) > (1/3)*BrickSideLengths(1)) & (Positions(3,1:UnitBrickNSpheres) < (2/3)*BrickSideLengths(1))); %Selecting all spheres in middle third of cube
light               % create a light
lighting gouraud    % preferred method for lighting curved surfaces
for count = 1:size(List,2)
    i = List(count);
    surf(Radii(i)*x_sph + Positions(1,i), Radii(i)*y_sph + Positions(2,i), Radii(i)*z_sph + Positions(3,i),'EdgeColor','none','FaceLighting','gouraud')
    
    if mod(count,1000)==0
        percent = 100*(count/size(List,2)); %Keeping track of progress
        disp(['Plotting middle third of Unit Cube ',num2str(percent), '% complete'])
    end
end
xlim([0, BrickSideLengths(1)]);
ylim([0, BrickSideLengths(1)]);
zlim([0, BrickSideLengths(1)]);
xlabel('$x, {\rm \mu m}$', 'Interpreter', 'latex');
ylabel('$y, {\rm \mu m}$', 'Interpreter', 'latex'); 
zlabel('$z, {\rm \mu m}$', 'Interpreter', 'latex');
view(60, 45);
hold off;

% Save figure - uncomment as needed
% savefig('Example1A_Method1_UnitCube_Middle.fig');
% print('Example1A_Method1_UnitCube_Middle','-dpdf');

%% Plot theoretical vs. actual distribution
FF=figure(14);
hold on
%plot histogram of obtained radii
histogram(Radii,30,'Normalization','pdf');

%plot Weibull PDF 
Weibull_scale = 31.4/2; % scale parameter lambda in Weibull distribution; in micrometers
Weibull_shape = 3.55; % shape parameter k in Weibull distribution; dimensionless
ProbabilityDistribution = makedist('Weibull','a',Weibull_scale ,'b', Weibull_shape);
X = linspace(0,max(Radii)*1.1);
Y = pdf(ProbabilityDistribution,X);
plot(X,Y, 'LineWidth', 3);
set(gca, 'FontSize', 14)
xlabel('Particle radius ($\mu$m)','Interpreter','latex')
ylabel('Probability density','Interpreter','latex')
legend('Actual', 'Weibull', 'Interpreter','latex', 'location', 'northeast')
xlim([0, max(Radii)*1.1]);
hold off

% Save figure - uncomment as needed
% savefig('Example1A_Method1_UnitCube_DistributionCurves.fig');
% print('Example1A_Method1_UnitCube_DistributionCurves','-dpdf');

%% Two Cubes joined to make a 2x1 brick
figure(21)
hold on
axis equal

%1:Number will be the spheres in the first unit cube, N is the total number
%of spheres, all cubes combined.
List = 1:2*UnitBrickNSpheres;% 
light               % create a light
lighting gouraud    % preferred method for lighting curved surfaces
for count = 1:size(List,2)
    i = List(count);
    surf(Radii(i)*x_sph + Positions(1,i), Radii(i)*y_sph + Positions(2,i), Radii(i)*z_sph + Positions(3,i),'EdgeColor','none','FaceLighting','gouraud')
    
    if mod(count,1000)==0
        percent = 100*(count/(2*UnitBrickNSpheres)); %Keeping track of progress
        disp(['Plotting 2 Cubes Joined ',num2str(percent), '% complete'])
    end
end
xlim([0, 2*BrickSideLengths(1)])
ylim([0, BrickSideLengths(1)])
zlim([0, BrickSideLengths(1)])
xlabel('$x, {\rm \mu m}$', 'Interpreter', 'latex');
ylabel('$y, {\rm \mu m}$', 'Interpreter', 'latex'); 
zlabel('$z, {\rm \mu m}$', 'Interpreter', 'latex');

%Overhead View
view(0,90);
hold off;

% Save figure - uncomment as needed
% savefig('Example1A_Method1_Overhead.fig');
% print('Example1A_Method1_TwoCubes_Overhead','-dpdf');


%% Total Build (in this example, it is 2x2 unit cubes)
figure(23)
hold on
axis equal

List = 1:FinalNSpheres;
light               % create a light
lighting gouraud    % preferred method for lighting curved surfaces
for count = 1:size(List,2)
    i = List(count);
    surf(Radii(i)*x_sph + Positions(1,i), Radii(i)*y_sph + Positions(2,i), Radii(i)*z_sph + Positions(3,i),'EdgeColor','none','FaceLighting','gouraud')
    
    if mod(count,1000)==0
        percent = 100*(count/FinalNSpheres); %Keeping track of progress
        disp(['Plotting total build ',num2str(percent), '% complete'])
    end
end
xlim([0, 2*BrickSideLengths(1)])
ylim([0, 2*BrickSideLengths(1)])
zlim([0, BrickSideLengths(1)])
xlabel('$x, {\rm \mu m}$', 'Interpreter', 'latex');
ylabel('$y, {\rm \mu m}$', 'Interpreter', 'latex'); 
zlabel('$z, {\rm \mu m}$', 'Interpreter', 'latex');
view(60, 30);
hold off;

% Save figure - uncomment as needed
% savefig('Example1A_Method1_TotalBuild.fig');
% print('Example1A_Method1_TotalBuild','-dpdf');
\end{lstlisting}

%-----------------
\section{Data generation and plotting script for sphere packing Method 2, Example 2 } \label{appendix:eg2:generate}

\begin{itemize}\setlength\itemsep{0.5ex}
  \item Main file: \verb|Example2_Method2_Generate_and_Plot.m|
  \item Output:
  \begin{itemize}\setlength\itemsep{0ex}
    \item Figure \ref{fig:Method2:all}.
    \item \verb|Matlab| data file \verb|Example2_Method2_Results.mat|
  \end{itemize}
  \item Required additional files (in the same folder):
  \begin{itemize}\setlength\itemsep{0ex}
    \item \verb|Method2GenerateSpheres.m|: the main Method 2 sphere filling routine
    \item Auxiliary routine files for \verb|Method2GenerateSpheres.m|:
    \begin{itemize}\setlength\itemsep{0ex}
      \item \verb|M2Position2Xmin.m|
      \item \verb|M2Position2Ymin.m|
      \item \verb|M2Position2Zmin.m|
      \item \verb|M2Position3.m|
      \item \verb|M2Search2D.m|
      \item \verb|M2Search3D.m|
    \end{itemize}
  \end{itemize}
\end{itemize}

\begin{lstlisting}[language=Matlab]
%% Matlab script for Example 2: Method 2, using the Gamma distribution
% Produces the sphere filling of the domain and plots of Figure 5
% Required files:
% * Method2GenerateSpheres.m -- main sphere filling routine
% * Auxiliary routine files for Method1GenerateSpheres.m :
%    - M2Position2Xmin.m
%    - M2Position2Ymin.m
%    - M2Position2Zmin.m
%    - M2Position3.m
%    - M2Search2D.m
%    - M2Search3D.m

clear all; close all; clc;

%% Sphere Packing parameters: Gamma distribution

rng(0); %random seed zero: Matlab default

Gamma_scale = 7; % scale parameter theta in the Gamma distribution; dimensional (micrometers)
Gamma_shape = 2;  % shape (k) k in Gamma distribution; dimensionless

%first parameter a: shape; 2nd parameter b: scale
ProbabilityDistribution = makedist('Gamma','a', Gamma_shape ,'b', Gamma_scale);
average_radius = mean(ProbabilityDistribution) %equal to Gamma_shape*Gamma_scale

FaceGoal = 1.0; 
BodyGoal = 0.9; 

%Unit brick: here unit cube
std_length = 30*average_radius; 
BrickSideLengths = [1;1;1] * std_length;

%numbers of cubes in x, y, z directions
BrickNumbers = [2;2;1];

SphereContactParameter = 0.1; 
ParentParameter = 0.5; 

%% Run the sphere generation Method 2 algorithm and save results

tic

[FinalNSpheres, UnitBrickNSpheres, Positions, Radii, Contacts, ListXmin, ListYmin, ListZmin, ListXmax, ListYmax, ListZmax] = ...
  Method2GenerateSpheres( ...
    ProbabilityDistribution, ...
    FaceGoal, BodyGoal, ...
    SphereContactParameter, ParentParameter, ...
    BrickSideLengths, BrickNumbers );

sphere_filling_total_time = toc

save('Example2_Method2_Gamma_Results.mat')

%% Method 2, plot desired vs. actual distribution: Fig.5c

FF=figure(14);
hold on;

%plot histogram of obtained radii
histogram(Radii,30,'Normalization','pdf');

%plot Weibull PDF 
X = linspace(0,max(Radii));
Y = pdf(ProbabilityDistribution,X);
plot(X,Y, 'LineWidth', 3);
set(gca, 'FontSize', 14)
xlabel('Particle radius','Interpreter','latex')
ylabel('Probability density','Interpreter','latex')
legend('Actual', 'Gamma', 'Interpreter','latex', 'location', 'northeast')
hold off

% savefig('Example2_Method2_Gamma_DistributionCurves.fig');
% print('Example2_Method2_Gamma_DistributionCurves','-dpdf');

%% Example 2, Method 2, Unit Brick Plot - open (Fig. 5a)
max_radius=max(Radii);
minimal_sphere_x=-max_radius;
maximal_sphere_x=max(Positions(1,:))+max_radius;
minimal_sphere_y=-max_radius;
maximal_sphere_y=max(Positions(2,:))+max_radius;
minimal_sphere_z=-max_radius;
maximal_sphere_z=max(Positions(3,:))+max_radius;

figure(11)
hold on
axis equal
[x_sph, y_sph, z_sph] = sphere;
percent = 0;
List = 1:UnitBrickNSpheres; %particles to display
light               % create a light
lighting gouraud    % preferred method for lighting curved surfaces
for count = 1:size(List,2)
    i = List(count);
    surf(Radii(i)*x_sph + Positions(1,i), Radii(i)*y_sph + Positions(2,i), Radii(i)*z_sph + Positions(3,i),'EdgeColor','none','FaceLighting','gouraud')
    
    if mod(count,1000)==0
        percent = 100*(count/UnitBrickNSpheres); %Keeping track of progress
        disp(['Plotting Open Unit Brick ',num2str(percent), '% complete'])
    end
    
end
xlim([0, BrickSideLengths(1)]);
ylim([0, BrickSideLengths(2)]);
zlim([0, BrickSideLengths(3)]);
xlabel('$x, {\rm \mu m}$', 'Interpreter', 'latex');
ylabel('$y, {\rm \mu m}$', 'Interpreter', 'latex'); 
zlabel('$z, {\rm \mu m}$', 'Interpreter', 'latex');
view(30, 30);
hold off;

% save as needed
% savefig('Example2_Method2_Gamma_UnitBrickOpen.fig');
% print('Example2_Method2_Gamma_UnitBrickOpen','-dpdf');

%% Example 2, Method 2, Unit Brick Plot - closed (Fig. 5b)
figure(12)
hold on
axis equal
[x_sph, y_sph, z_sph] = sphere;
percent = 0;
List = 1:UnitBrickNSpheres; %particles to display
light               % create a light
lighting gouraud    % preferred method for lighting curved surfaces
for count = 1:size(List,2)
    i = List(count);
    surf(Radii(i)*x_sph + Positions(1,i), Radii(i)*y_sph + Positions(2,i), Radii(i)*z_sph + Positions(3,i),'EdgeColor','none','FaceLighting','gouraud')
    
    if mod(count,1000)==0
        percent = 100*(count/UnitBrickNSpheres); %Keeping track of progress
        disp(['Plotting Closed Unit Brick ',num2str(percent), '% complete'])
    end
    
end

max_sphere_radius=1.05*max(Radii);
xlim([-max_sphere_radius, BrickSideLengths(1) + max_sphere_radius]);
ylim([-max_sphere_radius, BrickSideLengths(2) + max_sphere_radius]);
zlim([-max_sphere_radius, BrickSideLengths(3) + max_sphere_radius]);
xlabel('$x, {\rm \mu m}$', 'Interpreter', 'latex');
ylabel('$y, {\rm \mu m}$', 'Interpreter', 'latex'); 
zlabel('$z, {\rm \mu m}$', 'Interpreter', 'latex');
view(30, 30);
hold off;

% save as needed
% savefig('Example2_Method2_Gamma_UnitBrickClosed.fig');
% print('Example2_Method2_Gamma_UnitBrickClosed','-dpdf');

%% Example 2, Method 2, Total domain - top and bottom faces  (Fig. 5d)

figure(31)
hold on
axis equal
List = ListZmin';
List2 = ListZmax';
light               % create a light
lighting gouraud    % preferred method for lighting curved surfaces
for count = 1:size(List,2)
    i = List(count);
    surf(Radii(i)*x_sph + Positions(1,i), Radii(i)*y_sph + Positions(2,i), Radii(i)*z_sph + Positions(3,i),'EdgeColor','none','FaceLighting','gouraud')
end
xlim([minimal_sphere_x, maximal_sphere_x]);
ylim([minimal_sphere_y, maximal_sphere_y]);
zlim([minimal_sphere_z, maximal_sphere_z]);
xlabel('$x, {\rm \mu m}$', 'Interpreter', 'latex');
ylabel('$y, {\rm \mu m}$', 'Interpreter', 'latex'); 
zlabel('$z, {\rm \mu m}$', 'Interpreter', 'latex');

for count = 1:size(List2,2)
    i = List2(count);
    surf(Radii(i)*x_sph + Positions(1,i), Radii(i)*y_sph + Positions(2,i), Radii(i)*z_sph + Positions(3,i),'EdgeColor','none','FaceLighting','gouraud')
end
xlim([minimal_sphere_x, maximal_sphere_x]);
ylim([minimal_sphere_y, maximal_sphere_y]);
zlim([minimal_sphere_z, maximal_sphere_z]);
xlabel('$x, {\rm \mu m}$', 'Interpreter', 'latex');
ylabel('$y, {\rm \mu m}$', 'Interpreter', 'latex'); 
zlabel('$z, {\rm \mu m}$', 'Interpreter', 'latex');
view([30, 30]);
hold off;

% save as needed
% savefig('Example2_Method2_Gamma_UnitCube_SidesZ.fig');
% print('Example2_Method2_Gamma_UnitCube_SidesZ','-dpdf');

\end{lstlisting}

%-----------------
\section{Data generation and plotting script, Example 3} \label{appendix:eg3:generate}

\begin{itemize}\setlength\itemsep{0.5ex}
  \item Main file: \verb|Example3A_Cube_Hemisph_Generate_and_Plot.m|
  \item Input and output: see Section \ref{sec:eg3:hemisph}, Example 3A for details.
  \item Required additional files (in the same folder):
  \begin{itemize}\setlength\itemsep{0ex}
    \item \verb|Example3GenerateSpheres.m|: the main sphere filling routine
    \item Auxiliary routine files for \verb|Example3GenerateSpheres.m|:
    \begin{itemize}\setlength\itemsep{0ex}
      \item \verb|EG3Position2Ymin.m|
      \item \verb|EG3Position2Ymax.m|
      \item \verb|EG3Position2Zmin.m|
      \item \verb|EG3Position2Zmax.m|
      \item \verb|EG3Position3.m|
    \end{itemize}
  \end{itemize}
\end{itemize}

\begin{lstlisting}[language=Matlab]
%% Matlab script for Example 3: a brick (here cube) with partially spherical noundary

clear all; close all; clc;

% -- use Weibull distribution implemented in Matlab
% with parameters from Steuben, Iliopoulos, Michopoulos (2016)

rng(0); %random seed zero: Matlab default

Weibull_scale = 31.4/2; % scale parameter lambda for the radius in Weibull distribution; in micrometers. The scale parameter for *diameter* is 31.4.
Weibull_shape = 3.55; % shape parameter k in Weibull distribution; dimensionless

ProbabilityDistribution = makedist('Weibull','a',Weibull_scale ,'b', Weibull_shape);

average_radius = mean(ProbabilityDistribution);

FaceGoal = 0.4; 
BodyGoal = 0.4; 

% Unit brick: here a unit cube
std_length = 30*average_radius; 
BrickSideLengths = [1;1;1] * std_length;

% Radii of hemispheres defining the romain, located on x_min and x_max faces
% Recommended: (x-length of the domain)/2 or smaller
HemisphereRadii = [0.5; 0.5] * std_length;

SphereContactParameter = 0.1; 

tic;

[NSpheres, Positions, Radii, Contacts] = ...
  Example3GenerateSpheres( ...
    ProbabilityDistribution, ...
    BrickSideLengths, ...
    HemisphereRadii, ...
    FaceGoal, BodyGoal, ...
    SphereContactParameter ...
 );

sphere_filling_total_time = toc

%% Save all data 
save('Example3A_Cube_Hemisph_data.mat')

%% Plot the given Weibull disribution vs. the actual distribution
figure(1)
hold on
X = linspace(0,max(Radii));
Y = wblpdf(X,Weibull_scale,Weibull_shape);
histogram(Radii,30,'Normalization','pdf')
plot(X,Y, 'LineWidth', 3);
set(gca, 'FontSize', 16)
xlabel('Diameter of particle $\mu m$','Interpreter','latex')
ylabel('Frequency of Diameter','Interpreter','latex')
legend('Actual Distribution', 'Desired Dirstibution')
set(gcf,'color','w');

% save figure: uncomment if needed
% print('Radii_Distribution_SphericalBoundary','-dpdf');
% savefig('Radii_Distribution_SphericalBoundary.fig');                                        

%% Particle Placement Plot 
figure(2)
hold on
axis equal
[x, y, z] = sphere;
List = 1:NSpheres; 
light               
lighting gouraud    
for count = 1:size(List,2)
    i = List(count);
    surf(Radii(i)*x + Positions(1,i), Radii(i)*y + Positions(2,i), Radii(i)*z + Positions(3,i),'EdgeColor','none','FaceLighting','gouraud')
end
xlabel('$x, {\rm \mu m}$', 'Interpreter', 'latex');
ylabel('$y, {\rm \mu m}$', 'Interpreter', 'latex'); 
zlabel('$z, {\rm \mu m}$', 'Interpreter', 'latex');
set(gcf,'color','w');
view([45, 45])

% save figure: uncomment if needed
% print('ParticlePlacement_SphericalBoundary','-dpdf');
% savefig('ParticlePlacement_SphericalBoundary.fig');


%% First parents plot
x_max = BrickSideLengths(1); y_max = BrickSideLengths(2); z_max = BrickSideLengths(3);
x_min = 0; y_min = 0; z_min = 0;
x_max_input = BrickSideLengths(1); y_max_input = BrickSideLengths(2); z_max_input = BrickSideLengths(3);

figure(3)
hold on
axis equal
[x, y, z] = sphere;
List = 1:8; %These are the first parents

surf(HemisphereRadii(1)*x, HemisphereRadii(1)*y + y_max_input/2, HemisphereRadii(1)*z + z_max_input/2,'FaceAlpha',0.8); %First Spherical Boundary
surf(HemisphereRadii(2)*x + x_max_input, HemisphereRadii(2)*y + y_max_input/2, HemisphereRadii(2)*z + z_max_input/2,'FaceAlpha',0.8); %Second Spherical Boundary
light               
lighting gouraud    
for count = 1:size(List,2)
    i = List(count);
    surf(Radii(i)*x + Positions(1,i), Radii(i)*y + Positions(2,i), Radii(i)*z + Positions(3,i),'EdgeColor','none','Facecolor','r')
end
xlim([0 x_max])
ylim([0 y_max])
zlim([0 z_max])
plot3([x_min, x_max],[y_min, y_min],[z_min, z_min],'k')
plot3([x_min, x_max],[y_max, y_max],[z_min, z_min],'k')
plot3([x_min, x_min],[y_min, y_max],[z_min, z_min],'k')
plot3([x_max, x_max],[y_min, y_max],[z_min, z_min],'k')
plot3([x_min,x_min],[y_min,y_min],[z_min,z_max],'k')
plot3([x_max,x_max],[y_min,y_min],[z_min,z_max],'k')
plot3([x_max,x_max],[y_max,y_max],[z_min,z_max],'k')
plot3([x_min,x_min],[y_max,y_max],[z_min,z_max],'k')
plot3([x_min, x_max],[y_min, y_min],[z_max, z_max],'k')
plot3([x_min, x_max],[y_max, y_max],[z_max, z_max],'k')
plot3([x_min, x_min],[y_min, y_max],[z_max, z_max],'k')
plot3([x_max, x_max],[y_min, y_max],[z_max, z_max],'k')
xlabel('$x, {\rm \mu m}$', 'Interpreter', 'latex');
ylabel('$y, {\rm \mu m}$', 'Interpreter', 'latex'); 
zlabel('$z, {\rm \mu m}$', 'Interpreter', 'latex');
view([30, 30])

% save figure: uncomment if needed
% print('FirstParents_SphericalBoundary','-dpdf');
% savefig('FirstParents_SphericalBoundary.fig');

%% Plot the sphere filling and the domain boundaries
figure(4)
hold on
axis equal
[x, y, z] = sphere;
List = 1:NSpheres; 
light              
lighting gouraud   
for count = 1:size(List,2)
    i = List(count);
    surf(Radii(i)*x + Positions(1,i), Radii(i)*y + Positions(2,i), Radii(i)*z + Positions(3,i),'EdgeColor','none')
end
surf(HemisphereRadii(1)*x, HemisphereRadii(1)*y + y_max_input/2, HemisphereRadii(1)*z + z_max_input/2,'FaceAlpha',0.5,'Facecolor','r'); %First Spherical Boundary
surf(HemisphereRadii(2)*x + x_max_input, HemisphereRadii(2)*y + y_max_input/2, HemisphereRadii(2)*z + z_max_input/2,'FaceAlpha',0.5,'Facecolor','r'); %Second Spherical Boundary
xlim([0 x_max])
ylim([0 y_max])
zlim([0 z_max])
plot3([x_min, x_max],[y_min, y_min],[z_min, z_min],'k')
plot3([x_min, x_max],[y_max, y_max],[z_min, z_min],'k')
plot3([x_min, x_min],[y_min, y_max],[z_min, z_min],'k')
plot3([x_max, x_max],[y_min, y_max],[z_min, z_min],'k')
plot3([x_min,x_min],[y_min,y_min],[z_min,z_max],'k')
plot3([x_max,x_max],[y_min,y_min],[z_min,z_max],'k')
plot3([x_max,x_max],[y_max,y_max],[z_min,z_max],'k')
plot3([x_min,x_min],[y_max,y_max],[z_min,z_max],'k')
plot3([x_min, x_max],[y_min, y_min],[z_max, z_max],'k')
plot3([x_min, x_max],[y_max, y_max],[z_max, z_max],'k')
plot3([x_min, x_min],[y_min, y_max],[z_max, z_max],'k')
plot3([x_max, x_max],[y_min, y_max],[z_max, z_max],'k')
xlabel('$x, {\rm \mu m}$', 'Interpreter', 'latex');
ylabel('$y, {\rm \mu m}$', 'Interpreter', 'latex'); 
zlabel('$z, {\rm \mu m}$', 'Interpreter', 'latex');
view([30, 30])

% save figure: uncomment if needed
% print('ParticlesAndSphericalBoundaries_SphericalBoundary','-dpdf');
% savefig('ParticlesAndSphericalBoundaries_SphericalBoundary.fig');
\end{lstlisting}

\end{appendix}

\end{document}